%% file: main.tex
\documentclass{article}

\usepackage[preprint]{main}

\usepackage[utf8]{inputenc} 
\usepackage[T1]{fontenc}    
\usepackage{hyperref}       
\usepackage{url}            
\usepackage{booktabs}       
\usepackage{amsfonts}       
\usepackage{nicefrac}       
\usepackage{microtype}      
\usepackage[table,xcdraw]{xcolor}       
\usepackage{enumitem}       %
\usepackage{setspace} 
\usepackage{mathtools} 
\usepackage{algorithmic}

\usepackage{graphicx}
\usepackage{wrapfig}
\usepackage{pifont}
\usepackage{float} 
\usepackage{arydshln}  
\usepackage{bbding}
\usepackage{amssymb}
\usepackage{array}      
\usepackage{multirow}   
\usepackage{xcolor}     
\usepackage{colortbl}
\usepackage{adjustbox}
\usepackage{tablefootnote}
\usepackage{tabularx}
\usepackage[ruled,vlined]{algorithm2e}

\newtheorem{theorem}{Theorem}
\newtheorem{lemma}{Lemma}

\title{$\begin{array}{l}\includegraphics[height=2.6\fontcharht\font`\B]{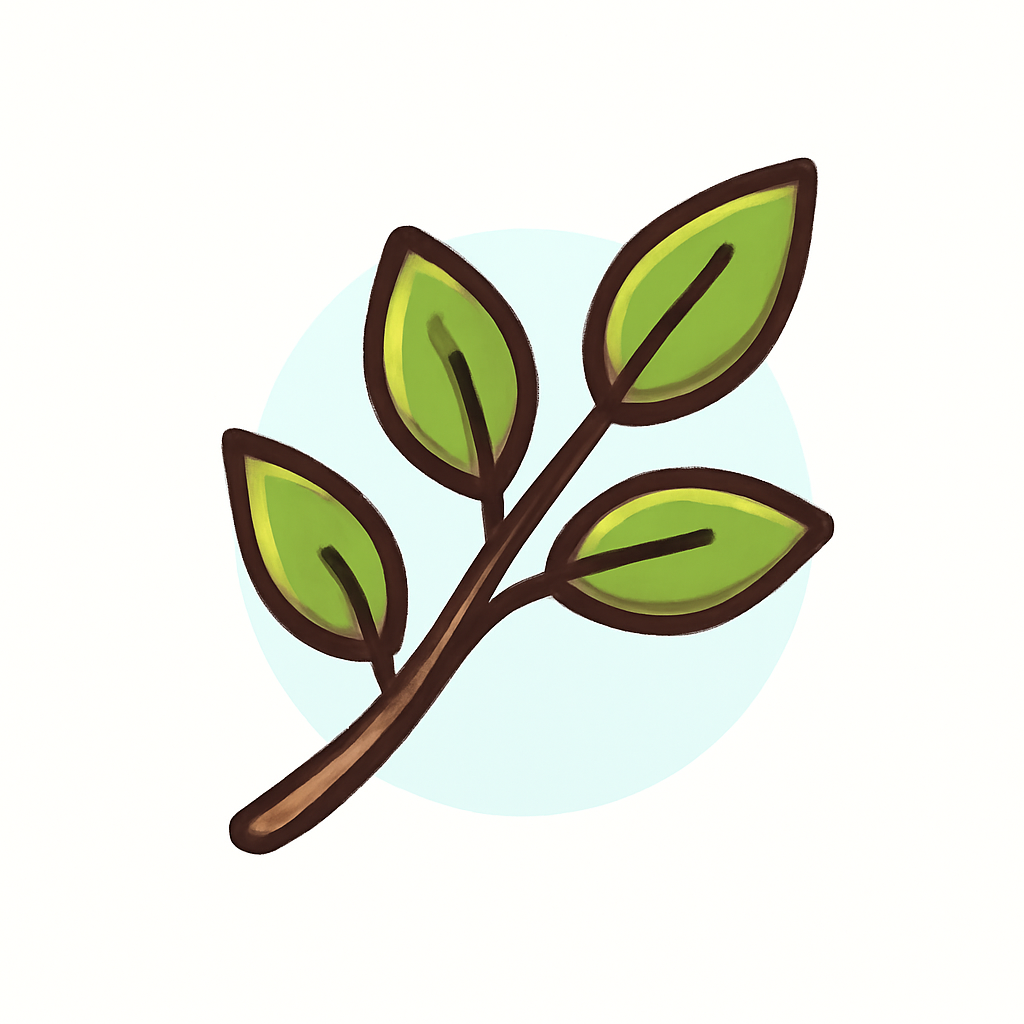}\end{array}$ Speculative Decoding via Hybrid Drafting and Rollback-Aware Branch Parallelism}

\vspace{-0.7in}
\author{
   Yuhao Shen\thanks{Equal contribution.}$^{\hspace{0.4em}\mathsection}$ \quad Junyi Shen\footnotemark[1]$^{\hspace{0.4em}\diamondsuit}$ \quad Quan Kong$^{\mathsection}$
  \\
  \textbf{Tianyu Liu}$^{\ddag}$
  \quad
  \textbf{Yao Lu}$^{\diamondsuit}$
  \quad
  \textbf{Cong Wang}\thanks{The Corresponding Authors.}$^{\hspace{0.4em}\mathsection}$
  \\
  ${}^\mathsection$Zhejiang University \quad 
  ${}^\diamondsuit$National University of Singapore \\
  ${}^\ddag$University of Science and Technology of China\\
    \texttt{\{riven, quankong, cwang85\}@zju.edu.cn}\\
    \texttt{\{j1shen, luyao\}@comp.nus.edu.sg}\\
    \texttt{tianyu\_liu@mail.ustc.edu.cn}\\
}

\begin{document}

\maketitle

\begin{abstract}
Speculative decoding (SD) has emerged as a promising technique to accelerate LLM inference by employing a small draft model to propose draft tokens in advance, and validating them in parallel with the large target model. However, the existing SD methods still remain constrained by their serialized execution, which causes the mutual waiting bubbles between the draft and target models. To address this challenge, we draw inspiration from branch prediction in modern processors and propose a novel framework \textbf{SpecBranch} to unlock branch parallelism in SD. Specifically, we first take an in-depth analysis of the potential of branch parallelism in SD, and recognize that the key challenge lies in the trade-offs between parallelization and token rollback. Based on the analysis, we introduce parallel speculative branches to preemptively hedge against likely rejections. Meanwhile, to enhance parallelism, we jointly orchestrate adaptive draft lengths with a hybrid combination of the implicit draft model confidence and explicit reusing of target model features. Extensive experiments across various models and benchmarks show that SpecBranch achieves over \textbf{1.8}$\times \sim$ \textbf{4.5}$\times$ speedups against the auto-regressive decoding and reduces rollback tokens by $\textbf{50}$\% for poorly aligned models, while maintaining an identical sampling distribution. Our code is available at \url{https://github.com/Sylvan820/Specbranch}.
\end{abstract}

\section{Introduction}


Recent advances in Large Language Models (LLMs), such as GPT-4~\cite{achiam2023gpt}, Gemini~\cite{team2023gemini}, Qwen3~\cite{bai2023qwen}, and DeepSeek~\cite{guo2025deepseek}, have revolutionized natural language processing~\cite{brown2020language}. However, their real-world deployment faces the critical challenge of inference latency due to auto-regressive token-by-token generation, which restricts LLMs to predicting one token at a time, creating a fundamental
bottleneck for real-time and large-scale applications.

\begin{figure*}[t]
\centering
\includegraphics[width=0.98\linewidth]{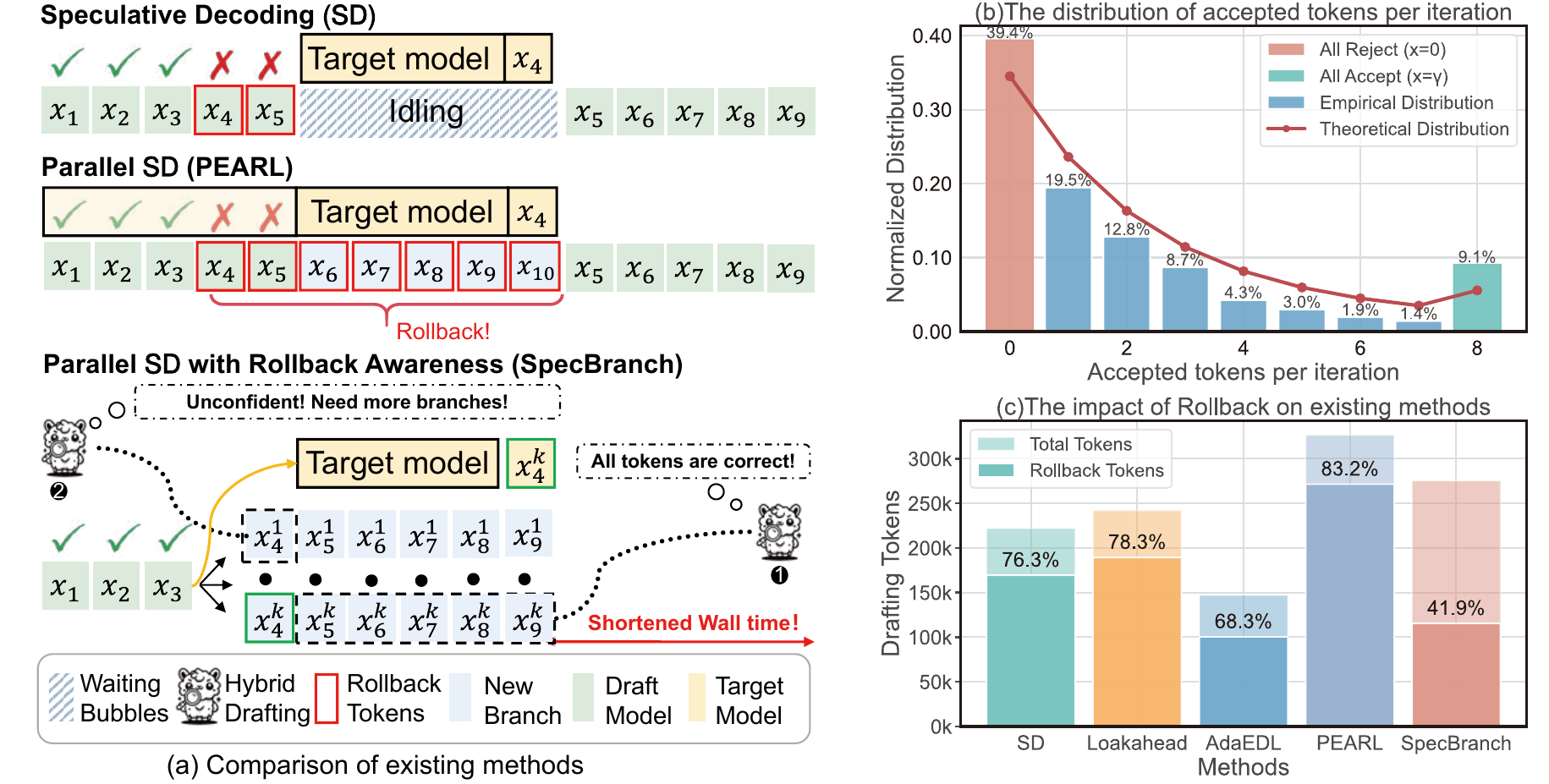}
\caption{{Architectural comparison and empirical analysis of SD frameworks: a) Vanilla SD, Parallel SD (PEARL), and Parallel SD with Rollback Awareness (SpecBranch). When rejection occurs at token $x_4$, PEARL's static pipeline forces verification of those ``doomed tokens'' $x_5-x_{10}$; SpecBranch dynamically terminates invalid branches and spawns new branches. b) Distribution of accepted tokens generally follows a truncated geometric distribution with longer token length (Vicuna 68M\&13B, $\gamma = 8$); c) Percentage of rollback tokens under different mechanisms.} }
\label{fig:1}
\vspace*{-0.2in}
\end{figure*}

To address this limitation, Speculative Decoding (SD) has emerged as a promising acceleration paradigm~\cite{leviathan2023fast, chen2023accelerating, stern2018blockwise, li2024eagle, li2024eagle2, liu2024parallel}. SD uses a small \emph{draft model} to proactively generate candidate tokens, which are then verified in parallel by the large \emph{target model}. By replacing serialized token generation with parallel validation, SD decouples the computational workload from sequence length. However, a critical serialization bottleneck still remains. As shown in Fig.~\ref{fig:1}(a), the draft and target models operate in strict alternation: the draft model idles during target model verification, but the target model cannot process new candidates until the draft model completes its proposal. This mutual dependency leads to \emph{pipeline bubbles}~\cite{narayanan2019pipedream} that neither model fully saturates the hardware resources.


Inspired by branch prediction in modern processors~\cite{jimenez2001dynamic,shi2019applying}, we allow the draft model to \textit{proactively generate speculative branches} concurrently with target model verification. Such parallel SD paradigm creates a two-stage pipeline in which the draft model token generation overlaps with the target model validation, effectively filling the inherent pipeline bubbles in vanilla SD. Prior works such as PEARL~\cite{liu2024parallel} use the target model to verify the first draft token during the drafting phase (pre-verify), and use the draft model to continue generating draft tokens during the verification phase (post-verify). However, unlike lockstep execution from the existing SD that discards tokens only with a local penalty (called ``rollback tokens'' henceforth), the paradigm shift to Parallel SD risks  \textbf{global invalidation} if a token is rejected which causes all subsequent tokens to be rejected and stall parallelism efficiency. This is exacerbated with longer token length as accepted tokens typically follow a truncated geometric distribution as shown in Fig.~\ref{fig:1}(b), creating a trade-off between parallelism and rollback. 


Unfortunately, PEARL inadequately addresses these challenges: \textbf{1) Pre-verify Rollback.} PEARL overlooks a critical condition for parallel acceleration: the tokens during the verification need to be \textbf{All-Accepted}; otherwise, PEARL degenerates to serialized execution and loses its parallel capacity. It verifies only the first token by the target model, while the system remains oblivious to mid-sequence rollback until the parallel verification completes (e.g., when $x_4$ is rejected in Fig.~\ref{fig:1}(b)). \textbf{2) Post-verify Rollback.} The static draft length lacks sufficient awareness of rollback and rejected tokens, which also undermine the benefits of parallelism (shown in Fig.~\ref{fig:1}(c) with a high percentage of rollback). Consequently, it leads to redundant computation of those ``doomed tokens'' and makes the target model to become a bottleneck for processing unnecessary tokens from invalidated branches, despite inevitable rollbacks. This is exacerbated in resource-constrained systems due to model misalignment from the parameter-imbalanced pairs (68M draft\&13B target). 

Although recent dynamic drafting methods, categorized as implicit (confidence-driven early stopping~\cite{li2024eagle2,liu2024kangaroo,agrawal2024adaedl,zhang2023draft}) or explicit (feature-based sequence modeling~\cite{zhang2024adaeagle}) partially mitigate the rollbacks, they face practical challenges from per-task threshold tuning, error compounding and low prediction accuracy. To this end, we propose \textbf{SpecBranch} with the following contributions:

\vspace*{-0.05in}
\begin{enumerate}[leftmargin=*,align=left]
    \item[\ding{71}] \textbf{Branch-Parallel Architecture}: We first establish theoretical models to quantify ideal parallel speculation and extend it to consider rollback penalties in practice. Guided by these insights, we propose a novel \emph{branch resampling} mechanism that introduces parallel speculative branches to preemptively hedge against likely rejections, while preserving the original sample distribution.
    \item[\ding{71}] \textbf{Hybrid Adaptive Drafting}: Based on extensive empirical analysis of adaptive draft structures, we are the first to unify the implicit (draft model confidence) and explicit (target model feature) methods into a hybrid framework that dynamically optimizes draft lengths. This effectively reduces the percentage of rollback and improves parallel efficiency.
    \item[\ding{71}] \textbf{Extensive Evaluation and Discussion}: We conduct extensive experiments across various models and tasks, demonstrating that SpecBranch consistently achieves a \textbf{1.8}$\times$to \textbf{4.5}$\times$ speedup without draft-model training and reduces rollback tokens by $\textbf{50}$\% for poorly aligned draft/target models.
\end{enumerate}


    
    
    
\section{Related Work}
\label{headings}
\textbf{Speculative Decoding}\quad 
While SD has demonstrated significant acceleration and lossless generalization, increasing the acceptance rate of draft tokens by the target model remains a critical challenge. Existing approaches rely on draft model training-based~\cite{cai2024medusa, li2024eagle, du2024glide} and training-free methods~\cite{fu2024break, chen2023accelerating, zhao2024ouroboros, liu2024parallel} to align the draft and target models. For instance, Medusa introduces auxiliary decoding heads to the target model~\cite{cai2024medusa}, while Eagle~\cite{li2024eagle} and Glide~\cite{du2024glide} reuse target model information to enhance token prediction accuracy. SpecInfer uses tree-based attention to efficiently verify multiple draft candidates in order to improve acceptance rates~\cite{chen2023accelerating}. On the other hand, training-free methods such as Lookahead decoding adopt a trajectory caching mechanism to store $n$-gram generation histories as draft candidates~\cite{fu2024break}. However, all these methods follow a sequential \emph{draft-then-verify} paradigm, which is fundamentally limited by the mutual waiting bottleneck. DSI~\cite{timor2024distributed} introduces a distributed speculation parallelism framework to orchestrate target and drafter instances that overlap in time. PEARL~\cite{liu2024parallel} introduces a parallel framework that verifies the first draft token while allowing the draft model to simultaneously generate additional tokens during verification. However, it overlooks the impact of rollback when verification fails, which negates the benefit of parallelism if not properly addressed.

\definecolor{mygreen}{RGB}{233,247,234}

\begin{table}[t]
\centering
\resizebox{\textwidth}{!}{
\begin{tabular}{l c c c c}
\toprule
\textbf{Methods} & \textbf{Parallel Drafting} & \textbf{Model-Training-free} & \textbf{Draft Structure Modeling} & \textbf{Speedup}\\ 
\midrule
Kangaroo~\cite{liu2024kangaroo}               & \textcolor{red}{\XSolidBrush} & \textcolor{red}{\XSolidBrush} & Implicit (Confidence) & - \\
EAGLE-2~\cite{li2024eagle2}                & \textcolor{red}{\XSolidBrush} & \textcolor{red}{\XSolidBrush} & Implicit (Confidence) & -\\
AdaEAGLE~\cite{zhang2024adaeagle}              & \textcolor{red}{\XSolidBrush} & \textcolor{red}{\XSolidBrush} & Explicit (Feature) & - \\
\hdashline
Lookahead Decoding~\cite{fu2024break}      & \textcolor{red}{\XSolidBrush} & \textcolor{green}{\Checkmark} & None & 1.1$\times$$\sim$1.8$\times$ \\
AdaEDL~\cite{agrawal2024adaedl}                 & \textcolor{red}{\XSolidBrush} & \textcolor{green}{\Checkmark} & Implicit (Entropy) & 1.4$\times$$\sim$3.0$\times$ \\
PEARL~\cite{liu2024parallel} & \textcolor{green}{\Checkmark}          & \textcolor{green}{\Checkmark} & None (Chunk-level) & 1.6$\times$$\sim$4.2$\times$  \\ 
\midrule
\rowcolor{mygreen}  
\textbf{SpecBranch (Ours)}       &  \textcolor{green}{\Checkmark} & \textcolor{green}{\Checkmark} & \textbf{Hybrid (Token-level)} & \textbf{1.8$\times$$\sim$4.5$\times$}  \\ 
\bottomrule
\end{tabular}
}
\caption{{Comparison of SpecBranch with the existing SD methods. SpecBranch is the first parallel framework with hybrid drafting structures that does not require additional training of draft models.}}  \label{table:related_works}
\vspace{-0.2in}
\end{table}

\textbf{Dynamic Drafting Structures}\quad 
Dynamic drafting structure is an effective approach to optimizing SD~\cite{li2024eagle2,liu2024kangaroo,agrawal2024adaedl,zhang2023draft,zhang2024adaeagle}. It adapts the draft sequence length or tree configuration (e.g., depth, width, shape) based on contextual speculation. Current methods to model drafting boundaries fall into two categories illustrated by Table~\ref {table:related_works}: \emph{implicit} and \emph{explicit}. Implicit methods rely on output distribution metrics (e.g., confidence, entropy) to dynamically terminate drafts~\cite{li2024eagle2,liu2024kangaroo,agrawal2024adaedl,zhang2023draft}. However, these require manually tuned thresholds and struggle to balance flexibility with computational overhead, particularly in trainable-head variants~\cite{huang2024specdec++,mamou2024dynamic}, which incur latency by predicting tokens individually. By eliminating per-token prediction, explicit methods such as AdaEAGLE~\cite{zhang2024adaeagle} directly estimate draft lengths based on target model features. While promising, direct estimation of long sequences introduces instability with low prediction accuracy. Essentially, none of these works addresses rollback, a key bottleneck where drafts waste tokens and stall the parallelism. This work proposes a hybrid framework to combine the implicit confidence-based termination with explicit sequence modeling that reduces rollback substantially and improves parallel efficiency.

\section{Preliminaries}    \label{sec:pre}

\textbf{Notations}\quad
We define the \emph{draft model} as $\mathit{M}_q$ and the \emph{target model} as $\mathit{M}_p$. Given a prefix $\mathbf{X}_{1:j} = (x_1, \cdots, x_j)$, $q(\cdot)$ and $p(\cdot)$ denote the probability distributions of the draft and target models, respectively. The speed ratio $c = T_p/T_q$ quantifies the relative latency. Token generation maps $\mathbf{X}_{1:j}$ to embeddings $E_{1:j}$, which transforms to latent features $F_{1:j}$ and generates the next-token distribution $p_{j+1}$ from the final feature vector $f_j$.



\textbf{Speculative Decoding}\quad
Speculative decoding accelerates autoregressive generation through parallel token verification. The draft model $\mathit{M}_q$ proposes $\gamma$ candidate tokens $\Tilde{\mathbf{X}}_{1:\gamma}$ with probabilities $\{q(x_{i}|\mathbf{X}_{1:i-1})\}_{i=1}^\gamma$. The target model $\mathit{M}_p$ computes true probabilities in one forward pass. The acceptance probability for each candidate $x_{i}$ is $\beta(t_{i}) = \min\left(1, \frac{p(t_{i}|\mathbf{X}_{1:i-1})}{q(t_{i}|\mathbf{X}_{1:i-1})}\right)$. We employ the $\text{Match}(p(t_i|\mathbf{X}_{1:i-1}), q(t_i|\mathbf{X}_{1:i-1}))$ function~\cite{zhao2024ouroboros} to represent the verification process, which identifies the set of tokens accepted following the verification step. If $x_{i}$ is rejected, subsequent candidates $\Tilde{\mathbf{X}}_{i+1:\gamma}$ are discarded, and a token is resampled from $\mathrm{norm}(\max(0, p(x_{i}) - q(x_{i})))$; if all $\gamma$ tokens are accepted, an additional token is sampled from $p(t_{\gamma+1})$~\cite{leviathan2023fast}.

\vspace{-0.05in}
\section{Analysis of Parallel Decoding}





\vspace{-0.05in}
\subsection{Theoretical Speedup}
\label{sec:analysis1}
\vspace{-0.05in}
We first quantify the theoretical speedup of parallel SD under different circumstances. The draft model generates $\gamma$ candidate tokens $\mathbf{X}_{1:\gamma}$, which are verified by the target model. Let $T_q = t$ denote the draft model's per-token generation time, and $T_p = ct$ be the target model's verification time. From~\cite{leviathan2023fast}, under full acceptance of $\gamma$ tokens, the baseline SD achieves $T_{\text{SD}} = \frac{\gamma \cdot T_q + T_p}{\gamma + 1} = \frac{\gamma + c}{\gamma + 1} \cdot t$.

\vspace{-0.05in}
\textbf{Parallel SD (Ideal)}\quad Under the ideal full acceptance condition, the theoretical per-token latency with parallel decoding (shown in Fig.~\ref{fig:1}(a)) can be derived as: 
\vspace{-0.05in}
\begin{equation}
    T_{\text{PSD}}=\frac{\max(\gamma t, ct)}{\gamma} = 
    \begin{cases} 
        t, & \gamma \geq c \\
        \frac{c}{\gamma}t, & \gamma < c 
    \end{cases}
    \label{eq:psd_ideal}
\end{equation}
Then the speedup ratio from SD is, $T_{\text{SD}}/T_{\text{PSD}} = \frac{\gamma + c}{\gamma+1}$ or $\frac{\gamma + c}{\gamma+1}\frac{c}{\gamma} $. When $\gamma \approx c$ and $c \gg 1$, PD achieves an optimal $2\times$ speedup against SD. For autoregressive decoding with $\gamma \approx c$, PD represents $c \times$ speedup. Nevertheless, the practical performance depends on the draft acceptance rate, which has been largely overlooked in the prior work~\cite{liu2024parallel}.

\begin{wrapfigure}{r}{0.35\textwidth}
    \vspace{-0.6cm}
    \begin{center}
    \includegraphics[width=0.35\columnwidth]{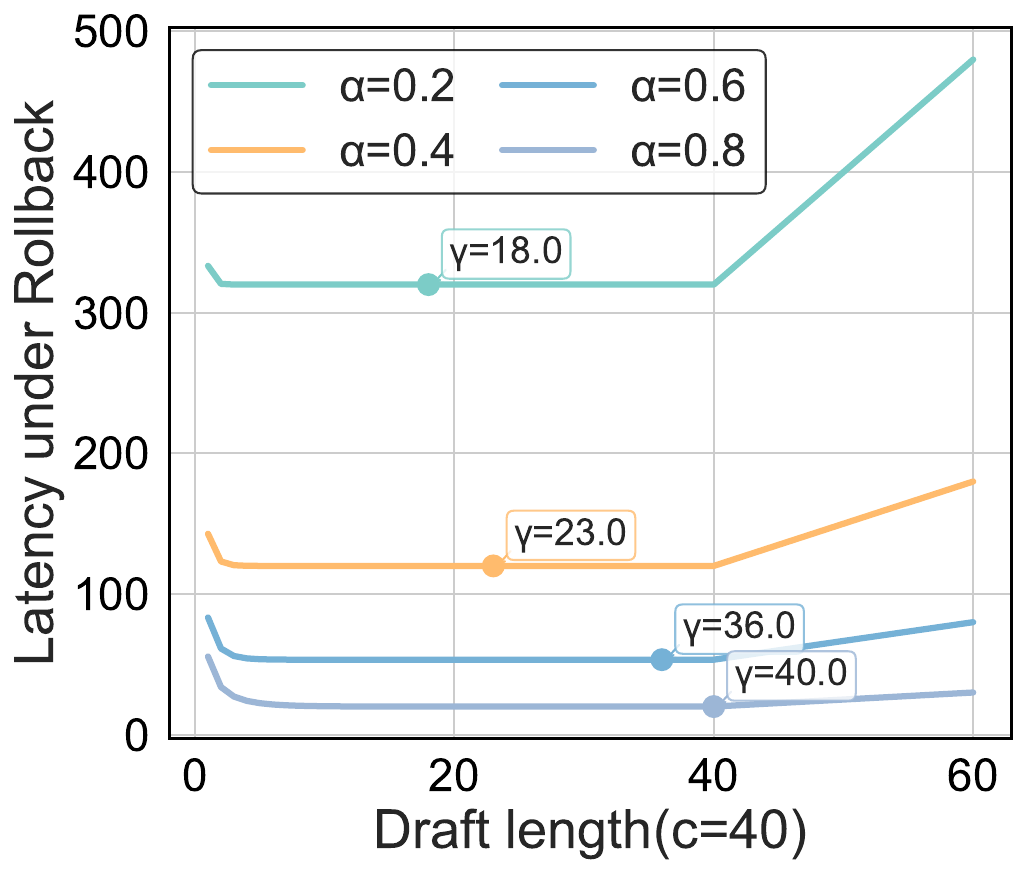}
    \vspace{-0.8cm}
    \caption{{Latency under rollback (Theorem 1). For different $\alpha$, the minimum values are presented (the curves have very mild slopes). }}
    \label{fig:theorem1}
    \end{center}
    \vspace{-0.21in}
\end{wrapfigure}
\textbf{Parallel SD (with Rollback)} \quad Recall $\beta$ as the acceptance rate and assume $\beta$s are i.i.d. with $\alpha = \mathbb{E}(\beta)$ as the expected acceptance rate (how well $M_q$ approximates $M_p$). When $k \leq \gamma$, the draft accepted length can be approximated by a truncated geometric distribution~\cite{leviathan2023fast} (shown in Fig.~\ref{fig:1}(b), detailed in Appendix~\ref{sec:appendix_result3}),
\begin{equation}
P(X = k) = (1 - \alpha) \cdot \alpha^k \cdot \mathbb{I}(k < \gamma) + \alpha^\gamma \cdot \mathbb{I}(k = \gamma)
\end{equation}
where $\alpha^{\gamma}$ is the probability of \textit{full acceptance} and $1-\alpha^{\gamma}$ is the probability of \textit{rollback}. The rollback penalty becomes severe when the draft model has limited capacity, which would cause the subsequent tokens to be discarded and revert parallelism back to serialized execution.  

\textbf{Theorem 1 (Latency under Rollback).} The per-token latency of parallel SD under rollback is,
\vspace{-0.15in}
\begin{equation}
T_{\text{PSD}_r}=\frac{2\cdot\max(\gamma t,ct)}{(1 +\alpha^\gamma)\cdot \frac{\alpha(1-\alpha^\gamma)}{1-\alpha}}=
\begin{cases}
\frac{2c t(1-\alpha)}{\alpha(1+\alpha^\gamma)(1 - \alpha^\gamma)}, & \gamma\leq c \\
\frac{2\gamma t(1-\alpha)}{\alpha(1+\alpha^\gamma)(1 - \alpha^\gamma)}, & \gamma>c 
\end{cases}
\label{eq:theorem1}
\end{equation}
We defer the proofs to Appendix~\ref{appendix:proof}. As visualized in Fig.~\ref{fig:theorem1}, the minimum latency occurs at the segment of $\gamma \leq c$, which theoretically validates the trade-off between parallelism and rollback: while small $\gamma$ underutilizes parallel resources, further increasing $\gamma$ beyond the minimum value leads to diminishing returns due to rollback accumulation. This trade-off is $\alpha$-dependent: for well-aligned models ($\alpha \rightarrow 1$), a larger $\gamma$ enjoys parallelism but for misaligned/capacity-constrained draft models $\alpha \leq 0.5$, the penalties from rollback dominate. Though insightful, this theoretical analysis only reflects the statistical properties rather than run-time dynamics. The actual accepted draft length is context-dependent and varies substantially across different iterations~\cite{zhang2024adaeagle} (detailed by Fig.~\ref{fig:appendix_length_change} in Appendix), which necessitates adaptive control of $\gamma$ rather than static configurations.




\begin{figure*}[t]
\centering
\includegraphics[width=1.00\linewidth]{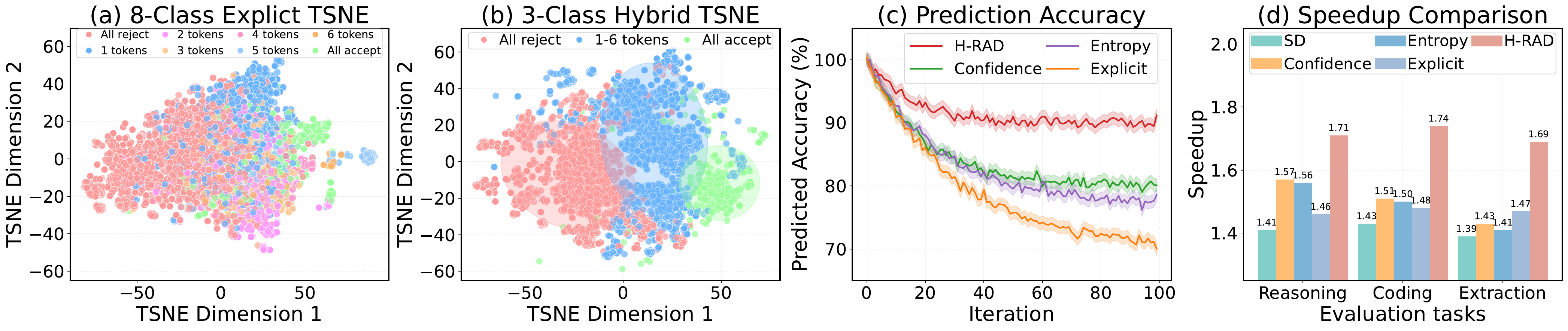}
\vspace*{-0.2in}
\caption{Empirical results of different drafting length estimation strategies: (a,b) comparison of T-SNE visualization for the explicit and the proposed hybrid methods from the MLP activations; (c) both implicit and explicit drafting structures have limitations of low prediction accuracy of the accepted draft length; (d) impact of different drafting schemes on acceleration potentials.}
\label{fig:motivation}
\vspace*{-0.15in}
\end{figure*}


\vspace{-0.05in}
\subsection{Analysis of Adaptive Draft Structures}  \label{sec:analysis2}
\vspace{-0.03in}
Thus, the next question is: ``How to optimize draft structures to balance parallelism and rollback?'' To answer this, we compare the implicit and explicit methods on LLaMA 68M and 7B across the MTbench datasets~\cite{zheng2023judging} for dialogue tasks. Implicit methods evaluate the confidence $\max_{x_{i}}q(x_{i})$~\cite{du2024glide} or entropy $1-\sqrt{\lambda H(x_{i})}$~\cite{agrawal2024adaedl} (positively correlated with acceptance rate) against pre-determined thresholds $\epsilon$, but the optimal $\epsilon$ are difficult to find across different tasks, models, temperatures (detailed in Appendix~\ref{sec:appendix_result6}). Additionally, token-level predictions would cause the error-compounding effect with higher instability across different tasks. On the other hand, explicit methods like AdaEAGLE~\cite{zhang2024adaeagle} predict accepted length $\gamma$ directly using target model features. Unfortunately, the discriminative power of the explicit methods also declines with the increase of accepted length as seen by the visually inseparable clusters in Fig.~\ref{fig:motivation}(a). As a result, the explicit method results in lower prediction accuracy than the implicit method as indicated in Fig.~\ref{fig:motivation}(c), despite less overall variance of the ultimate speedup across different tasks (Fig.~\ref{fig:motivation}(d)). A partial reason for this difficulty is due to the imbalanced geometric distribution of accepted lengths and the limited contextual information from a single feature layer adopted by~\cite{zhang2024adaeagle}. Based on these insights, we introduce hybrid rollback-aware branch parallelism.



 


\section{SpecBranch: Hybrid Drafting and Rollback-Aware Branch Parallelism}
\label{sec:design}
To address the dual challenges of pipeline bubbles and amplification of rollback cost in SD, we present SpecBranch with two novel components: 1) \emph{Hybrid Rollback-Aware Draft Structures (H-RAD)}, which combine the draft model confidence early stopping and target model feature reuse for adaptive draft lengths; 2) \emph{Branch Resampling}, a parallel drafting-verification mechanism that eliminates sequential bottlenecks through context-aware parallelism. Meanwhile, we provide a detailed step-by-step profiling example in Appendix~\ref{sec:appendix_example} for better understanding.

\subsection{H-RAD: Hybrid Rollback-aware Draft Structure} 
\label{sec:h-rad}
\begin{figure*}[t]
\centering
\includegraphics[width=1.00\linewidth]{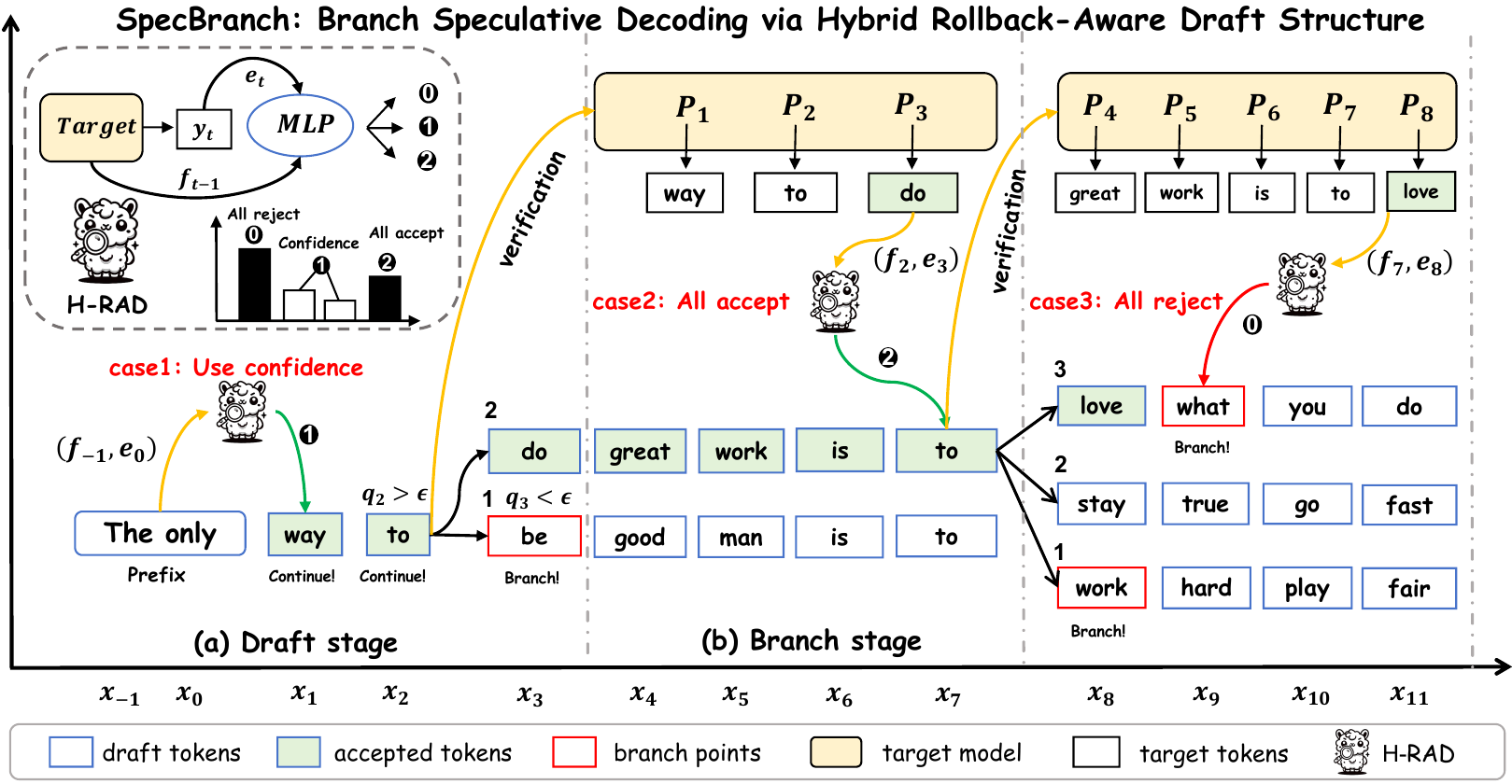}
\vspace*{-0.2in}
\caption{Architecture of SpecBranch. \textbf{Case 1 (Use Confidence)}: H-RAD outputs $s_{t} = 1$, indicating the branch point is determined by the draft model's confidence. \textbf{Case 2 (All Accept)}: H-RAD outputs $s_{t} = 2$, predicting that all tokens should be retained, and the branch point is the next round's first token, \textit{`work'}. \textbf{Case 3 (All Reject)}: H-RAD outputs $s_{t} = 0$, predicting that no tokens should be retained, and the branch point is the first token of this round, \textit{`what'}.}
\label{fig:framework}
\vspace*{-0.15in}
\end{figure*}

As illustrated by Fig.~\ref{fig:framework} (Case 1), H-RAD predicts the optimal draft lengths before branch resampling in the draft stage. Unlike PEARL's limited pre-verification of only the first token by target model~\cite{liu2024parallel}, we leverage the insights from the truncated geometric distribution of accepted tokens in Fig.~\ref{fig:1}(b) to build a hybrid and lightweight predictor to reduce rollback.

\textbf{Hybrid Drafting Length Prediction}\quad 
Given a prefix, our goal is to accurately predict the draft length $\gamma \leq c$. Since direct regression or multi-class classification of $\gamma$ suffers from low accuracy, the proposed hybrid design reduces the $\gamma$-class classification into a $3$-class classification problem. This is inspired by an intriguing bimodal phenomenon that target model features from multiple layers exhibit strong separability for the \emph{fully accepted} and \emph{rejected} cases, whereas the intermediate cases can be resolved by the implicit approach. As illustrated in Fig.~\ref{fig:motivation}(b), the hybrid method provides a more separable clustering compared to the overlapping distributions. Thus, we learn a lightweight MLP,
\begin{eqnarray}
&& \mathbf{z}_t = \text{Concat}({f}_{t-1}, \mathbf{e}_t) = \text{Concat}\left( \mathbf{h}_{t-1}^1, \cdots, \mathbf{h}_{t-1}^K, \mathbf{e}_t \right) \in \mathbb{R}^{K \cdot L \cdot D_{\text{layer}} + D_{\text{emb}}} \label{eq:z_t} \\
&& s_t = \arg\max(\text{Softmax}(\text{MLP}(\mathbf{z}_t))) \in \{0, \, 1, \, 2\}, \label{eq:s_t}
\end{eqnarray}
where we extract $K$ hidden states ${h}_{t-1}$ from the target model’s last $K$ layers and concatenate them with the new token embedding $e_t$ to $({f}_{t-1}, \mathbf{e}_t)$. To capture richer context than single-layer approaches~\cite{li2024eagle, zhang2024adaeagle}, our method uses multiple layers for better length prediction. The output $s_t$ initiates a new hybrid drafting strategy $\mathcal{H}_t$ with three classes,
\begin{equation}
\mathcal{H}_t = 
\begin{cases} 
\emptyset & \text{if } s_t = 0 \quad (\text{Hard signal: All Reject}), \\
\left\{ x \in \mathbf{X}_{1:\gamma} \mid q(x) > \epsilon\right\} & \text{if } s_t = 1 \quad (\text{Soft signal: Confidence}), \\
\mathbf{X}_{1:\gamma} & \text{if } s_t = 2 \quad (\text{Hard signal: All Accept}).
\end{cases}
\end{equation}
$\mathcal{H}_t$ yields hybrid decisions of $s_0$ and $s_2$ as \emph{hard signals}, and uses draft model confidence $q(x)$ for the intermediate \emph{soft signal} $s_1$. Here, the hard and soft signals refer to direct and pending decisions, respectively. According to the empirical distribution, most tokens are handled by hard signals at once (all accept or reject), while a small fraction is verified by soft signals later. Such a hybrid approach improves the prediction accuracy of the explicit method as well as reduces the compounding errors from the implicit methods. H-RAD predicts the draft length and branch points, providing a dynamic decision for branch resampling, as described in Section~\ref{sec:branch_resampling}. 


\subsection{Branch Resampling: Parallel Drafting During Verification} \label{sec:branch_resampling}
 
\textbf{Branch Resampling} \quad H-RAD identifies unconfident tokens as branch points through $\mathcal{H}_t$ in three cases as shown in Fig.~\ref{fig:framework}. Then it splits the draft sequence at branch point $x_b$ and sends the prefix $\mathbf{X}_{1:b-1}$ to the target model to verify. Meanwhile, it also spawns $k$ parallel branches via Top-$k$ resampling from the draft model's confidence distribution $q(x_b)$. For example, as illustrated by Fig.~\ref{fig:framework} case 1, in the sequence \textit{`The only way to be'}, H-RAD outputs $s_{t} = 1$ to use confidence and the token \textit{`be'} ($q_3 < \epsilon$) triggers a new branch \textit{`do'} with \textit{`way to'} to be verified by the target model,     
\begin{equation}
\mathcal{B} = \text{TopK}\left( q(x_b), \, k \right) = \left\{ x_b^1, x_b^2, \dots, x_b^k \right\}, \hspace{0.05in} \text{where} \hspace{0.05in} k = \max\left( 1, \, \left\lfloor k_{\text{max}} \cdot (1 - q(x_b)) \right\rfloor \right),
\end{equation}
in which $k$ is adaptively controlled and scales inversely with $x_b$'s confidence $q(x_b)$. Lower confidence in $x_b$ indicates a lower acceptance rate~\cite{du2024glide}, thus spawning more branches to help SD hedge against likely rejections. Each branch $x_b^i \in \mathcal{B}$ generates subsequent tokens independently using the shared KV-Cache of the prefix $\mathbf{X}_{1:b-1}$ and branch token $x_b^i$ to avoid redundant computation.:
\begin{equation}
\mathbf{X}_{\text{branch}}^i = \mathbf{X}_{1:b-1} \oplus x_b^i \oplus M_q \left( \text{KV-Cache}(\mathbf{X}_{1:b-1} \oplus x_b^i) \right).
\end{equation}
The maximum draft length per branch is constrained by the draft/target model speed ratio $c$, ensuring simultaneous verification and drafting to eliminate pipeline bubbles.

\textbf{Branch Verification} \quad 
During token drafting, the target model $M_p$ concurrently verifies the last round tokens $\mathbf{X}_{1:b-1}$ using $\text{Match}(p(x_i|\mathbf{X}_{1:i-1}), q(x_i|\mathbf{X}_{1:i-1})) $ (Section~\ref{sec:pre}). If any token $x_i$ is rejected, all subsequent tokens are discarded, and the target model resamples a new token, back to the draft stage. Unlike vanilla SD, if all tokens in $\mathbf{X}_{1:b-1}$ are accepted, we do not need to resample a new token. Instead, we verify the branch point $x_b$ using $p(x_b)$ derived from $x_{b-1}$, apply branch speculative sampling algorithm to sample or adjust the distribution (details in Algorithm ~\ref{alg:bss}), consistent with ~\citep{li2024eagle,miao2024specinfer}, ensuring that the distribution of the output branch point aligns with the target LLM. The branch point verification is formalized as:
\begin{equation}
\mathcal{V} = \text{Match}\bigl( \underbrace{\bigl( q(x_b^1) \ldots, q(x_b^k) \bigr)}_{\text{Draft probability}} , \underbrace{\bigl( p(x_b^1), \ldots, p(x_b^k) \bigr)}_{\text{Target probability}} \bigr).
\end{equation}
Here, $\mathcal{V}$ represents the accepted branch point or resampled token after verification. If $\exists$ $x_b^i \in \mathcal{V}$, we then remain the corresponding branch, discarding all non-selected branches and their associated KV-Cache. Unlike tree-based methods that spawn branches at every token (expensive KV-Cache growth)~\citep{miao2024specinfer}, SpecBranch limits branching to uncertainty points identified by H-RAD and avoids the need for complex tree attention verification. We provide more details in Appendix~\ref{sec:appendix_future2}.

\vspace{-0.03in}
\textbf{Posterior Drafting in the Branch Stage} \quad \label{sec:Posterior Drafting}
As shown in Fig.~\ref{fig:framework} (Case 2/3), the parallel branch stage introduces a potential temporal mismatch between drafting and verification. During the drafting stage, H-RAD predicts the accepted length based on the previous features \textit{before the draft model generates tokens}. However, in the branch stage, the tokens from the previous round have not been verified yet, which means H-RAD cannot immediately access those reliable target model features with sufficient guidance from the history. To address this mismatch, we propose a posterior approach for selecting the retained tokens \textit{after the branch generates tokens}. Once parallel verification is completed, for the remaining $\mathcal{V}$ branch, we use $(f_{t-1}, e_t)$ from the current round as input to H-RAD, and select $\mathcal{H}_t$ after the verification step. This ensures that tokens for the next round are selected based on the most up-to-date context, effectively resolving the mismatch as detailed in Appendix~\ref{sec:appendix_future2}.


\begin{table}[t]
\centering
\vspace{-2pt}
\setlength{\tabcolsep}{4pt}
\renewcommand{\arraystretch}{0.7} 
\tiny 
\begin{adjustbox}{width=\textwidth}
\begin{tabular}{>{\centering\arraybackslash}p{1.2cm} p{1.2cm} cccccc c c}
\toprule
\multirow{2}{*}{\textbf{Models}} & \multirow{2}{*}{\textbf{Methods}} & \multicolumn{2}{c}{\textbf{HumanEval}} & \multicolumn{2}{c}{\textbf{GSM8K}} & \multicolumn{2}{c}{\textbf{CNN/DM}} & \textbf{Speed} & \textbf{Avg.} \\
\cmidrule(lr){3-4} \cmidrule(lr){5-6} \cmidrule(lr){7-8}
& & M & Speedup & M & Speedup & M & Speedup & \textbf{(tokens/s)} & \textbf{Speedup} \\
\midrule
\multirow{5}{*}{\centering \begin{tabular}{@{}c@{}} LLaMA\\ 68M\&7B \end{tabular}}
                  & SpS          & 2.64 & 1.46$\times$ & 3.32 & 1.74$\times$ & 2.26 & 1.42$\times$ & 63.04 & 1.54$\times$ \\
                  & AdaEDL         & 2.49 & 1.54$\times$ & 3.25 & 1.89$\times$ & 2.19 & 1.46$\times$ & 67.21 & 1.63$\times$ \\
                  & Lookahead       & 1.48 & 1.31$\times$ & 1.96 & 1.71$\times$ & 1.45 & 1.25$\times$ & 57.63 & 1.42$\times$ \\
                  & PEARL       & 2.79 & 1.69$\times$ & 3.82 & 1.86$\times$ & 2.64 & 1.66$\times$ & 71.22 & 1.74$\times$ \\
                  & \cellcolor{mygreen}\cellcolor{mygreen}\textbf{SpecBranch} & \cellcolor{mygreen}\textbf{3.24} & \cellcolor{mygreen}\textbf{2.04}$\times$ & \cellcolor{mygreen}\textbf{4.46} & \cellcolor{mygreen}\textbf{2.12}$\times$ & \cellcolor{mygreen}\textbf{3.17} & \cellcolor{mygreen}\textbf{1.87}$\times$ & \cellcolor{mygreen}\textbf{82.41} & \cellcolor{mygreen}\textbf{2.01}$\times$ \\
\midrule
\multirow{5}{*}{\centering \begin{tabular}{@{}c@{}} Vicuna\\ 68M\&13B \end{tabular}}
                  & SpS          & 2.87 & 1.79$\times$ & 2.54 & 1.56$\times$ & 2.07 & 1.45$\times$ & 48.79 & 1.60$\times$ \\
                  & AdaEDL         & 2.77 & 1.95$\times$ & 2.46 & 1.68$\times$ & 2.01 & 1.53$\times$ & 51.84 & 1.72$\times$ \\
                  & Lookahead        & 1.76 & 1.57$\times$ & 1.83 & 1.59$\times$ & 1.52 & 1.23$\times$ & 43.95 & 1.46$\times$ \\
                  & PEARL        & 3.11 & 2.02$\times$ & 2.83 & 1.61$\times$ & 2.89 & 1.68$\times$ & 53.31 & 1.77$\times$ \\
                  & \cellcolor{mygreen} \textbf{SpecBranch} & \cellcolor{mygreen}\textbf{3.69} & \cellcolor{mygreen}\textbf{2.47}$\times$ & \cellcolor{mygreen}\textbf{3.29} & \cellcolor{mygreen}\textbf{1.95}$\times$ &\cellcolor{mygreen} \textbf{3.21} & \cellcolor{mygreen}\textbf{1.89}$\times$ & \cellcolor{mygreen}\textbf{62.57} & \cellcolor{mygreen}\textbf{2.10}$\times$ \\
\midrule
\multirow{5}{*}{\centering \begin{tabular}{@{}c@{}} Deepseek\\ 1.3B\&33B \end{tabular}}
                  & SpS          & 4.45 & 2.16$\times$ & 3.85 & 1.86$\times$ & 3.91 & 1.96$\times$ & 31.38 & 1.99$\times$ \\
                  & AdaEDL         & 4.12 &2.35$\times$ & 3.57 & 2.01$\times$ & 3.74 & 2.16$\times$ & 34.22 & 2.17$\times$ \\
                  & Lookahead        & 2.36 & 1.77$\times$ & 1.74 & 1.43$\times$ & 1.89 & 1.65$\times$ & 25.55 & 1.62$\times$ \\
                  & PEARL        & 16.97 & 3.39$\times$ & 8.28 & 2.78$\times$ & 6.45 & 2.63$\times$ & 46.17 & 2.93$\times$ \\
                  & \cellcolor{mygreen} \textbf{SpecBranch} & \cellcolor{mygreen} \textbf{22.52} & \cellcolor{mygreen}\textbf{3.71}$\times$ & \cellcolor{mygreen}\textbf{10.19} & \cellcolor{mygreen}\textbf{3.02}$\times$ & \cellcolor{mygreen}\textbf{7.96} & \cellcolor{mygreen}\textbf{2.97}$\times$ & \cellcolor{mygreen}\textbf{50.94} & \cellcolor{mygreen}\textbf{3.23}$\times$ \\
\midrule
\multirow{5}{*}{\centering \begin{tabular}{@{}c@{}} LLaMA-3.1\\8B\&70B \end{tabular}}
                  & SpS          & 5.25 & 2.41$\times$ & 5.15 & 2.31$\times$ & 5.09 & 2.11$\times$ & 16.21 & 2.28$\times$ \\
                  & AdaEDL         & 4.96 & 2.55$\times$ & 4.97 & 2.37$\times$ & 4.85 & 2.19$\times$ & 16.91 & 2.37$\times$ \\
                  & Lookahead        &-&-&-&- &-&-&-&-\\
                  & PEARL        & 17.28 & 3.75$\times$ & 14.33 & 3.35$\times$ & 7.51 & 3.04$\times$ & 24.03 & 3.38$\times$ \\
                  & \cellcolor{mygreen}\textbf{SpecBranch} & \cellcolor{mygreen}\textbf{21.74} & \cellcolor{mygreen}\textbf{4.02}$\times$ & \cellcolor{mygreen}\textbf{18.08} & \cellcolor{mygreen}\textbf{3.67}$\times$ & \cellcolor{mygreen}\textbf{9.41} & \cellcolor{mygreen}\textbf{3.37}$\times$ & \cellcolor{mygreen}\textbf{26.27} & \cellcolor{mygreen}\textbf{3.69}$\times$ \\
\bottomrule
\end{tabular}
\end{adjustbox}
\caption{{Comparison with existing baselines on HumanEval~\cite{chen2021codex}, GSM8K~\cite{Cobbe_Kosaraju_Bavarian_Hilton_Nakano_Hesse_Schulman_2021} and CNN/DM~\cite{Nallapati_Zhou_dos} (mean accepted length $M$, speedup ratio, and speed). ``--'' indicate incompatibility: baseline implementations (transformers $=4.36.2$) conflict with LLaMA 3.1’s dependency on $\geq 4.43.0$. }   }
\vspace*{-0.3in}
\label{tab:mainresult1}
\end{table}
\vspace{-3pt}

\section{Experiments}
\label{sec:Experiments}
\vspace{-0.05in}
\textbf{Implementation Details}\quad
We evaluate the effectiveness of SpecBranch across diverse LLM configurations, particularly focusing on
scenarios where draft models have significantly fewer parameters (68M) and weak alignment with target models (7B-70B), and speedup ratios $c \in [4,15]$ ($c$ is rounded up to the integer value). This includes LLaMA~\cite{miao2024specinfer, touvron2023llama} (68M, 7B, $c = 10$), Vicuna~\cite{yang2024multi, zheng2023judging} (68M, 13B, $c = 15$), models with better alignment such as Deepseek-Coder~\cite{deepseek-coder} (1.3B, 33B, $c = 4$) and LLaMA-3.1~\cite{grattafiori2024llama3herdmodels} (8B, 70B, $c = 5$). We assess SpecBranch across several text generation tasks, including HumanEval (code generation)~\cite{chen2021codex}, GSM8K (arithmetic reasoning)~\cite{Cobbe_Kosaraju_Bavarian_Hilton_Nakano_Hesse_Schulman_2021}, CNN/DM (summarization)~\cite{Nallapati_Zhou_dos}, and Spec-Bench (widely-adopted six sub-tasks)~\cite{xia2024unlocking}. All of our experiments are conducted on NVIDIA A100-PCIE-40G GPUs. More details are provided in Appendix~\ref{sec:appendix_eval}.

\textbf{Baseline Methods}\quad We evaluate against the vanilla autoregressive decoding ($1.00\times$ baseline) and $4$ model training-free methods. \textbf{(1) Speculative Decoding (SpS)}~\cite{chen2023accelerating}: Standard implementation where a draft model generates $\gamma$ tokens for parallel verification. \textbf{(2) AdaEDL}~\cite{agrawal2024adaedl}: Early-stopping via entropy-based thresholds to terminate low-probability drafts. \textbf{(3) Lookahead Decoding}~\cite{fu2024break}: Token-level speculation using cached $n$-gram matches without auxiliary draft models. \textbf{(4) PEARL}~\cite{liu2024parallel}: Pipeline parallelism with pre/post-verification to overlap draft-target model execution.

\textbf{Evaluation Metrics}\quad
We report widely used metrics for SD: Mean Accepted Length $M$, Wall-Time Speedup Ratio, and Speed (tokens/sec). In SpecBranch, $M$ represents the continuously accepted length~\cite{liu2024parallel}. We also introduce a new metric, Rollback Rate (RB), defined as \(RB = \frac{\# \text{Rollback tokens}}{\# \text{Total tokens}},\) which quantifies computational waste from invalid drafts. Here, Rollback tokens count the tokens discarded after verification in the draft forward time. More details can be found in Appendix~\ref{sec:appendix_eval3}.

\textbf{H-RAD Training}\quad 
The training data for H-RAD pairs the feature vector $z_t$ from Eq.\eqref{eq:z_t} with the corresponding three-class labels $s_t$. We implement a lightweight three-layer MLP with ReLU activation. Training is performed \emph{offline} for $20$ epochs and $32$ batch size using the AdamW~\cite{loshchilov2017decoupled} optimizer. The training converges within $5$ minutes on a single NVIDIA A100 GPU and eliminates the need for costly online fine-tuning while maintaining adaptability to diverse text domains.



\begin{table*}[t]
  \centering
  \vspace{-2pt}
  \resizebox{\textwidth}{!}{
    \begin{tabular}{c@{\hspace{0.2cm}}l@{\hspace{0.2cm}}c@{\hspace{0.2cm}}c@{\hspace{0.2cm}}c@{\hspace{0.2cm}}c@{\hspace{0.2cm}}c@{\hspace{0.2cm}}c@{\hspace{0.2cm}}c@{\hspace{0.2cm}}c@{\hspace{0.2cm}}c@{\hspace{0.2cm}}c@{\hspace{0.2cm}}c@{\hspace{0.2cm}}c@{\hspace{0.2cm}}c}
    \toprule
    \multirow{2}[4]{*}{\textbf{Models}} & \multirow{2}[4]{*}{\textbf{Methods}} &   \multicolumn{2}{c}{\textbf{MT Bench}}   &   \multicolumn{2}{c}{\textbf{QA}} &  \multicolumn{2}{c}{\textbf{Sum}}  &  \multicolumn{2}{c}{\textbf{Math}} & \multicolumn{2}{c}{\textbf{RAG}} & \multicolumn{2}{c}{\textbf{Trans}} &   \multirow{2}[4]{*}{\textbf{Avg.}} \\
\cmidrule(lr){3-4}  \cmidrule(lr){5-6} \cmidrule(lr){7-8} \cmidrule(lr){9-10} \cmidrule(lr){11-12} \cmidrule(lr){13-14}         &         & M & Speedup & M & Speedup  & M & Speedup & M & Speedup & M & Speedup & M & Speedup &  \\
    \midrule
        \multirow{5}{*}{\large Vicuna} & SpS  &  2.63   &    1.74$\times$      &   2.47    &   1.64$\times$ & 2.58 &  1.70$\times$  & 2.37 & 1.55$\times$ &  2.47 &  1.56$\times$  & 2.57&  1.65$\times$ & 1.64$\times$ \\
          & AdaEDL  &  2.50  &  {1.80$\times$}  &   2.43   &  1.67$\times$ & 2.64 &  1.72$\times$  & 2.31 & 1.62$\times$ &  2.21 &  1.57$\times$  & 2.45 &  1.75$\times$ & 1.69$\times$ \\ 
          & Lookahead&  1.56   &  1.31$\times$    &   1.41   &   1.23$\times$ & 1.49 &  1.25$\times$  & 1.71 & 1.46$\times$ &  1.38 &  1.15$\times$  & 1.32&  1.10$\times$ & 1.25$\times$ \\ 
          & PEARL  &  2.62  &  {1.78$\times$}  &   2.45   &  1.64$\times$ & \textbf{2.78} &  \textbf{1.83}$\times$  & 2.63 & 1.67$\times$ &  2.61 &  1.66$\times$  & 2.89 &  2.05$\times$ & 1.77$\times$ \\
          &  \cellcolor{mygreen}\textbf{SpecBranch} &  \cellcolor{mygreen}\textbf{3.11}  & \cellcolor{mygreen}\textbf{2.09}$\times$  &   \cellcolor{mygreen}\textbf{2.67}  &   {\cellcolor{mygreen}\textbf{1.83}$\times$}& \cellcolor{mygreen}2.72 &  \cellcolor{mygreen}1.78$\times$  & \cellcolor{mygreen}\textbf{2.86} & \cellcolor{mygreen}\textbf{1.89$\times$} &  \cellcolor{mygreen}\textbf{2.83} &  \cellcolor{mygreen}\textbf{1.86$\times$}  & \cellcolor{mygreen}\textbf{3.32} &  \cellcolor{mygreen}\textbf{2.30$\times$} & \cellcolor{mygreen}\textbf{1.96$\times$} \\
    \midrule
    \multirow{5}{*}{\large LLaMA-3.1} & SpS &  4.67  &   2.31$\times$    &     4.57       &  2.27$\times$ & 5.09 & 1.98$\times$ & 5.01 & 2.44$\times$& 5.08 & 2.02$\times$ & 5.52 & 2.57$\times$ & 2.27$\times$\\
          & AdaEDL  &   4.31   &  2.43$\times$     &  4.23   & 2.30$\times$ &  4.83 & 2.05$\times$ & 4.94 & 2.46$\times$ & 4.86 & 2.13$\times$ & 5.24 & 2.65$\times$ & 2.34$\times$ \\ 
          & Lookahead &  -  &  - &   -   &  - & - &  - & - & - &  - &  -  & - &  - & - \\ 
          & PEARL &  8.46  &  {2.96$\times$}  &   8.37   &  3.27$\times$ & 9.10&  3.32$\times$  & 12.53 & 3.39$\times$ &  8.35 &  \textbf{3.41}$\times$  & 12.59 &  4.22$\times$ & 3.43$\times$ \\
        & \cellcolor{mygreen}\textbf{SpecBranch}&\cellcolor{mygreen}\textbf{10.85} & \cellcolor{mygreen}\textbf{3.24\(\times\)} &\cellcolor{mygreen}\textbf{10.59} & {\cellcolor{mygreen}\textbf{3.45\(\times\)}} & \cellcolor{mygreen}\textbf{11.40} & \cellcolor{mygreen}\textbf{3.63\(\times\)} & \cellcolor{mygreen}\textbf{15.76} & \cellcolor{mygreen}\textbf{3.78\(\times\)} & \cellcolor{mygreen}\textbf{9.16} & \cellcolor{mygreen}3.40\(\times\) & \cellcolor{mygreen}\textbf{16.64} & \cellcolor{mygreen}\textbf{4.51\(\times\)} & \cellcolor{mygreen}\textbf{3.67\(\times\)} \\
    \bottomrule
    \end{tabular}%
    }
\vspace*{-0.1in}
\caption{Comparison with the existing baselines on Spec-Bench~\cite{xia2024unlocking}. }
\label{tab:mainresult2}
\vspace*{-0.15in}
\end{table*}
\subsection{Main Results}



\begin{wrapfigure}{r}{0.35\textwidth}
    \vspace{-0.6cm}
    \begin{center}
    \includegraphics[width=0.35\columnwidth]{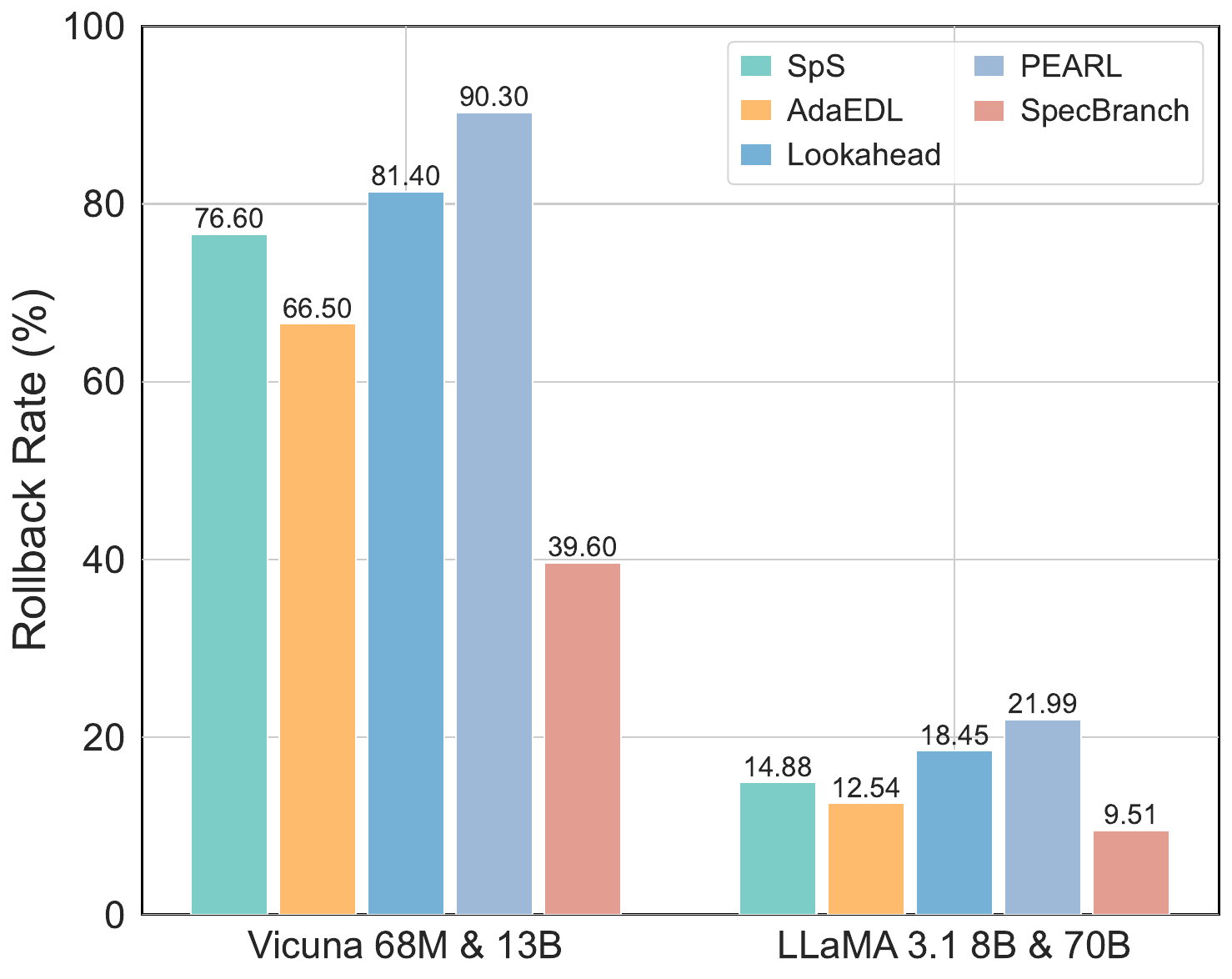}
    \vspace{-0.6cm}
    \caption{{Comparison of Rollback Rates on HumanEval.}}
    \label{fig:rollback}
    \end{center}
    \vspace{-0.17in}
\end{wrapfigure}

The main results from Tables~\ref{tab:mainresult1}, ~\ref{tab:mainresult2} and Fig.~\ref{fig:rollback} are explained below: \textbf{(\uppercase\expandafter{\romannumeral1})} On HumanEval, GSM8K, and CNN/DM benchmarks, SpecBranch shows superior efficiency over prior methods, achieving consistent speedups of $1.9 \times$ to $4.0 \times$ over vanilla autoregressive decoding. On Spec-Bench, our method also achieves significant acceleration from $1.8 \times$ to $4.5 \times$ on six diverse sub-tasks, indicating its robustness and versatility. \textbf{(\uppercase\expandafter{\romannumeral2})} For poorly aligned models with small $\alpha$ (LLaMA, Vicuna), rollback dominates parallel acceleration. SpecBranch improves over PEARL by 15\% due to rollback awareness and H-RAD’s hybrid design. As shown in Fig.~\ref{fig:rollback}, SpecBranch reduces rollback from $66$-$90$\% to under $40$\%, yielding longer generated lengths $M$ compared to the baseline (Tables~\ref{tab:mainresult1}, ~\ref{tab:mainresult2}). This validates the capability of H-RAD to terminate those draft paths with ultimate failure. \textbf{(\uppercase\expandafter{\romannumeral3})} For well-aligned models (Deepseek, LLaMA-3.1), SpecBranch improves parallelism and resource utilization through branch resampling. E.g., it achieves $3.23\times$ speedup vs. PEARL’s $2.93\times$ (10.2\%), with a $4.14\times$ improvement of the average accepted length $M$ against SpS on code generation tasks. It also reduces the rollback rate by $10$\% and improves speedup by $8$\% compared to PEARL. These results empirically validate the trade-off in Theorem 1 and more details are provided in Appendix~\ref{sec:appendix_results}.



\subsection{Ablation Studies}
\label{sec:ablation}
\begin{figure*}[h]
\vspace*{-0.1in}
\centering
\includegraphics[width=1.00\linewidth]{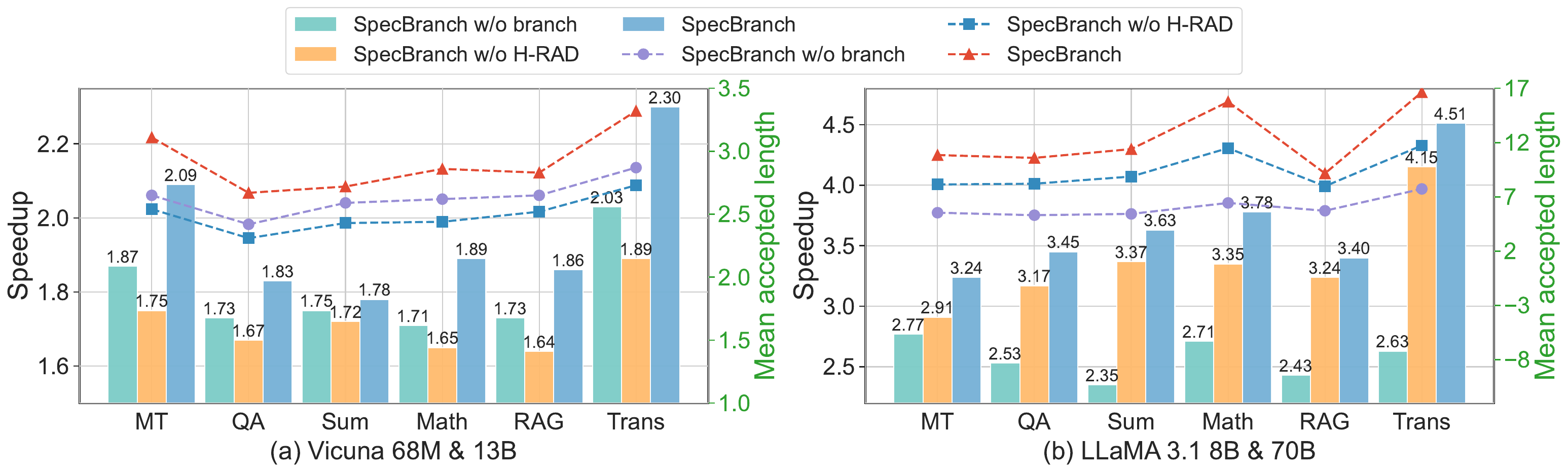}
\vspace*{-0.2in}
\caption{Component Ablation on the Spec-Bench benchmark: (a) for poorly aligned model pairs like Vicuna 68M-13B, H-RAD provides higher contributions; (b) for better-aligned models such as LLaMA-3.1 8B-70B, branch resampling plays a more important role in acceleration.}
\vspace*{-0.1in}
\label{fig:ablation}
\vspace*{-0.05in}
\end{figure*}

\textbf{Component Analysis}\quad Our ablation studies first isolate the two core components by removing: (1) branch resampling (SpecBranch \textit{w/o branch}), or (2) H-RAD (SpecBranch \textit{w/o H-RAD}). Fig.~\ref{fig:ablation} reveals that the absence of either component degrades performance. H-RAD dominates efficiency gains for misaligned model pairs like Vicuna 68M-13B, which improves the speedup from $1.72\times$ to $1.95\times$. On the other hand, branch resampling contributes more with better-aligned pairs like LLaMA-3.1 (8B-70B). Both components offer complementary advantages to reduce rollback and improve parallelism by adapting to different model capacity. Meanwhile, branch parallelism and H-RAD are both modular designs, which are orthogonal to methods like EAGLE~\cite{li2024eagle,li2024eagle2} and can be quickly adapted, and we leave for future exploration. We provide more discussions in Appendix~\ref{sec:appendix_future2}


\textbf{Hyperparameter Sensitivity}\quad Since H-RAD still entails the confidence threshold from the implicit methods, we evaluate its sensitivity to such threshold and feature layer hyperparameters in Tables~\ref{tab:thresholds} and~\ref{tab:feature} by integrating H-RAD with vanilla SD. Table~\ref{tab:thresholds} shows that while speed for implicit methods drops from $64$ to $49$ tokens/s as $\epsilon$ increases, H-RAD only decreases from $72$ to $67$ tokens/s, with less dependence on the threshold. The explicit variant in Table~\ref{tab:feature} reveals diminishing returns with more feature layers ($K$), that increasing context layers $K$ from $4$ to $32$ has marginal gains of $1-2$ tokens/s but $8\times$ more memory overhead. Thus, we choose $K = 4$ to balance the speed and memory. We provide more results in Appendix~\ref{sec:appendix_result6}.

\begin{table}[h]
\centering
\begin{minipage}{0.48\columnwidth}  
\centering
\resizebox{\columnwidth}{!}{
\begin{tabular}{cccc}
\toprule
\multirow{1}{*}{\textbf{$\epsilon$}} & \textbf{{Implict(Confidence)}} & \textbf{Implict(Entropy)} & \textbf{Hybrid(H-RAD)} \\
\midrule
0.1 & 61.05 & 60.28 & 70.32 \\
\rowcolor{mygreen}\textbf{0.2} & \textbf{64.26} & \textbf{63.03} & \textbf{72.15} \\
0.4 & 61.12 & 59.21 & 71.02 \\
0.6 & 56.43 & 54.73 & 69.91 \\
0.8 & 53.21 & 52.29 & 68.62 \\
0.9 & 49.46 & 48.18 & 67.31 \\
\bottomrule
\end{tabular}
}
\caption{{Results of Stop thresholds $\epsilon$ of LLaMA\\ 68M\&7B on HumanEval. (tokens/sec)}} 
\label{tab:thresholds}  
\end{minipage}%
\hspace{0.02\columnwidth}  
\begin{minipage}{0.48\columnwidth}  
\centering  
\resizebox{\columnwidth}{!}{
\begin{tabular}{cccc}
\toprule
\multirow{1}{*}{\textbf{$K$}} & \textbf{{HumanEval(coding)}} & \textbf{GSM8K(reason)} & \textbf{CNN/DM(sum)} \\
\midrule
1 & 62.35 & 63.28 & 52.82 \\
2 & 64.06 & 76.14 & 57.46 \\
\rowcolor{mygreen}\textbf{4} & \textbf{72.15} & \textbf{79.24} & \textbf{63.46} \\
8 & 73.28 & 80.22 & 63.86 \\
16 & 73.83 & 81.27 & 64.35 \\
32 & 74.18 & 81.33 & 64.41 \\
\bottomrule
\end{tabular}
}
\caption{{Results of feature layers $K$ of LLaMA 68M\&7B on H-RAD+SD. (tokens/sec)}} \label{tab:feature}  
\end{minipage}
\vspace*{-0.15in}
\end{table}

\begin{figure*}[h]
\centering
\includegraphics[width=1.00\linewidth]{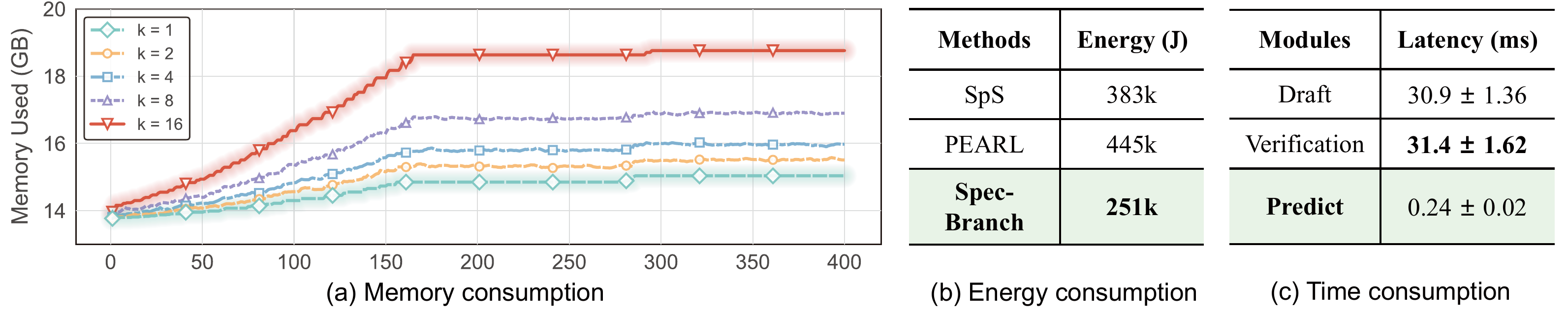}
\vspace*{-0.2in}
\caption{Resource consumption of SpecBranch through \texttt{NVIDIA DCGM}. (a) Trace of memory footprint of SpecBranch inference with different number of branches $k$ on HumanEval using LLaMA-3.1 8B-70B; (b)(c) Energy and time cost on HumanEval using Vicuna 68M-13B.}
\vspace*{-0.1in}
\label{fig:consumption}
\end{figure*}

\textbf{Resource Consumption}\quad We further validate SpecBranch's effectiveness across three dimensions of resource consumption. For memory consumption, we test the largest LLaMA-3.1 8B-70B models on HumanEval; for energy/time cost, we use poorly-aligned Vicuna 68M-13B on HumanEval. Fig.~\ref{fig:consumption}(a) shows that, due to the shared prefix KV-Cache, memory consumption for parallel branches with varying $k$ (from $1$ to $16$) only has a slight increment to 28\% of the baseline model parameters. In SpecBranch, $k$ is dynamically adjusted based on the draft model’s confidence at the branch point with $k_{\text{max}}$ typically capped by $6$. Unlike tree-based approaches~\cite{miao2024specinfer}, SpecBranch strategically spawns sparse branch points at high-impact tokens, which improves computational and memory efficiency by avoiding unnecessary branching. By the rollback-aware designs, Fig.~\ref{fig:consumption}(b) demonstrates that SpecBranch reduces energy consumption by 43\% against PEARL. Fig.~\ref{fig:consumption}(c) shows the time cost for each SpecBranch module, where the H-RAD prediction cost is negligible (0.38\% of total latency). SpecBranch eliminates the mutual waiting bubbles between draft and target models with almost identical execution time ($30.9$ vs $31.4$ ms per step). We provide more details in Appendix~\ref{sec:appendix_result5} ~\ref{sec:appendix_memory}.


\vspace{-0.05in}
\section{Conclusion}
We propose SpecBranch, a rollback-aware framework that accelerates auto-regressive decoding for LLMs. By enabling concurrent branch drafting and verification, SpecBranch addresses the bottleneck of serialized execution via two key innovations of branch resampling and hybrid drafting structures, which significantly improve acceleration and efficiency. Experiments show that SpecBranch achieves $1.8 \sim 4.5\times$ speedup and reduces computational waste up to 50\% for poorly aligned models, which enhances the efficiency in real-world applications. Although SpecBranch significantly reduces energy consumption compared to PEARL, it may still lead to higher resource consumption compared to auto-regressive decoding, which may pose a challenge for real-world deployment.

\section*{Acknowledgements}
This work is supported by Zhejiang Provincial National Science Foundation of China under Grant No. LZ25F020007 and National Science Foundation of China under grants 62576310, 62394341.


\setlength{\bibsep}{6pt} 
\bibliographystyle{plain}
\let\cleardoublepage\clearpage
\bibliography{reference}

\appendix
\section{Procedures of SpecBranch}
\input{pseudo}

\section{Proof of Theorem 1}  \label{appendix:proof}

We re-state Theorem 1 (Section~\ref{sec:analysis1}) here for convenience with the basic assumption from~\cite{leviathan2023fast} that the token acceptance is i.i.d. with $\mathbb{P}(\text{accept}) = \alpha$. Recall that the generation time $T_q = t$, the verification time $T_p = ct$, and any rollback would trigger a full retry of $\gamma$ tokens. To prove Theorem 1, we present Lemma 1 on the expected token draft length first. 
\begin{lemma}[Expected Draft Accepted length] \label{lem:expectation}
For truncated geometric distribution, \\$X \sim \text{TruncGeo}(\alpha, \gamma)$,
\begin{equation}
E[X] = \frac{\alpha(1 - \alpha^{\gamma})}{1 - \alpha}    
\end{equation}
\end{lemma}
\textit{Proof.}
\[
E[X] = \sum_{k=0}^\gamma k \cdot \mathbb{P}(X=k) = \sum_{k=0}^{\gamma-1} k \cdot (1-\alpha)\alpha^{k} + \gamma\cdot\alpha^{\gamma}
\]
Let $S = \sum_{k=0}^{\gamma-1} \alpha^k = \frac{1 - \alpha^{\gamma}}{1 - \alpha}$. Take differentiation regarding $\alpha$, we have,
\[
\frac{dS}{d\alpha} = \sum_{k=0}^{\gamma-1} k \alpha^{k-1} = \frac{1 - \gamma\alpha^{\gamma-1}+ (\gamma-1)\alpha^{\gamma}}{(1 - \alpha)^2}
\]
\[
E[X] = (1-\alpha)\alpha \cdot \frac{dS}{d\alpha} + \gamma\cdot\alpha^{\gamma} = \frac{\alpha(1 - \alpha^{\gamma})}{1 - \alpha} \quad 
\]

\begin{theorem}[Latency under Rollback]
\label{thm:bdr}
The latency of parallel SD under rollback is,
\[
T_{\text{BD}_r} = \frac{2 \cdot \max(\gamma t, ct)}{(1 + \alpha^{\gamma})\cdot \frac{\alpha(1 - \alpha^{\gamma})}{1 - \alpha}} 
= 
\begin{cases} 
\displaystyle \frac{2ct(1 - \alpha)}{\alpha(1+\alpha^\gamma)(1 - \alpha^\gamma)}, & \gamma \leq c \\
\displaystyle \frac{2\gamma t(1 - \alpha)}{\alpha(1+\alpha^\gamma)(1 - \alpha^\gamma)}, & \gamma > c 
\end{cases}
\]
\end{theorem}

\textit{Proof.} Define the acceptance vector $\mathbf{\omega} = (\omega_1, \dots, \omega_\gamma) \in \{0, 1\}^\gamma$, where $\omega_i = 1$ if and only if token $i$ is accepted. The accepted token count is,
\[
X = \sum_{i=1}^\gamma \omega_i, \quad \mathbb{P}(\omega_i = 1) = \alpha \ (\text{i.i.d.}).
\]
To compute the total number of tokens with retry, define two rounds of: 1) $\gamma$ tokens (accepted if $\mathbf{\omega} = \mathbf{1}$); 2) Retry if Round 1 fails, which yields $E[X]$ tokens. Thus, the total expectation is
\[
E_{\text{total}} = \alpha^\gamma (\gamma + E[X]) + (1 - \alpha^\gamma) \frac{(E[X] - \gamma\alpha^\gamma)}{1 - \alpha^\gamma}  = (1+\alpha^\gamma)\cdot E[X] \tag{1}
\]
This implies that Parallel SD (with Rollback) achieves an acceleration factor of $(1 + \alpha^\gamma) \times$ compared to the vanilla SD (with Rollback). As $\alpha$ approaches 1, the acceleration ratio reaches $2 \times$, matching the acceleration of the Ideal Parallel SD in Eq. \eqref{eq:psd_ideal}. Thus, the total time for the two rounds (parallel generation/verification):
\[
T_{\text{total}} = 2 \cdot \max(\gamma t, ct) \tag{2}
\]
\[
T_{\text{PSD}_r} = \frac{T_{\text{total}}}{E_{\text{total}}} = \frac{2 \cdot \max(\gamma t, ct)}{(1+\alpha^\gamma) \cdot \frac{\alpha(1 - \alpha^\gamma)}{1 - \alpha}} \tag{3}
\]
For different cases of length $\gamma$, we have,

\underline{Case 1: $\gamma \leq c$} (Verification time $ct > \gamma t$):
\[
T_{\text{PSD}_r} = \frac{2ct}{(1+\alpha^\gamma) \cdot \frac{\alpha(1 - \alpha^\gamma)}{1 - \alpha}} = \frac{2ct(1-\alpha)}{\alpha(1+\alpha^\gamma)(1 - \alpha^\gamma)}.
\]
\underline{Case 2: $\gamma > c$} (Generation time $\gamma t > ct$):
\[
T_{\text{PSD}_r} = \frac{2\gamma t}{(1+\alpha^\gamma) \cdot \frac{\alpha(1 - \alpha^\gamma)}{1 - \alpha}} = \frac{2\gamma t(1-\alpha)}{\alpha(1+\alpha^\gamma)(1 - \alpha^\gamma)}.
\]
(Simplified using $\gamma t = c t \cdot \frac{\gamma}{c} > ct$). This completes the proof of Theorem \ref{thm:bdr}.


\section{Profiling Example}\label{sec:appendix_example}
\begin{figure*}[h]
\vspace*{-0.1in}
\centering
\includegraphics[width=1\linewidth]{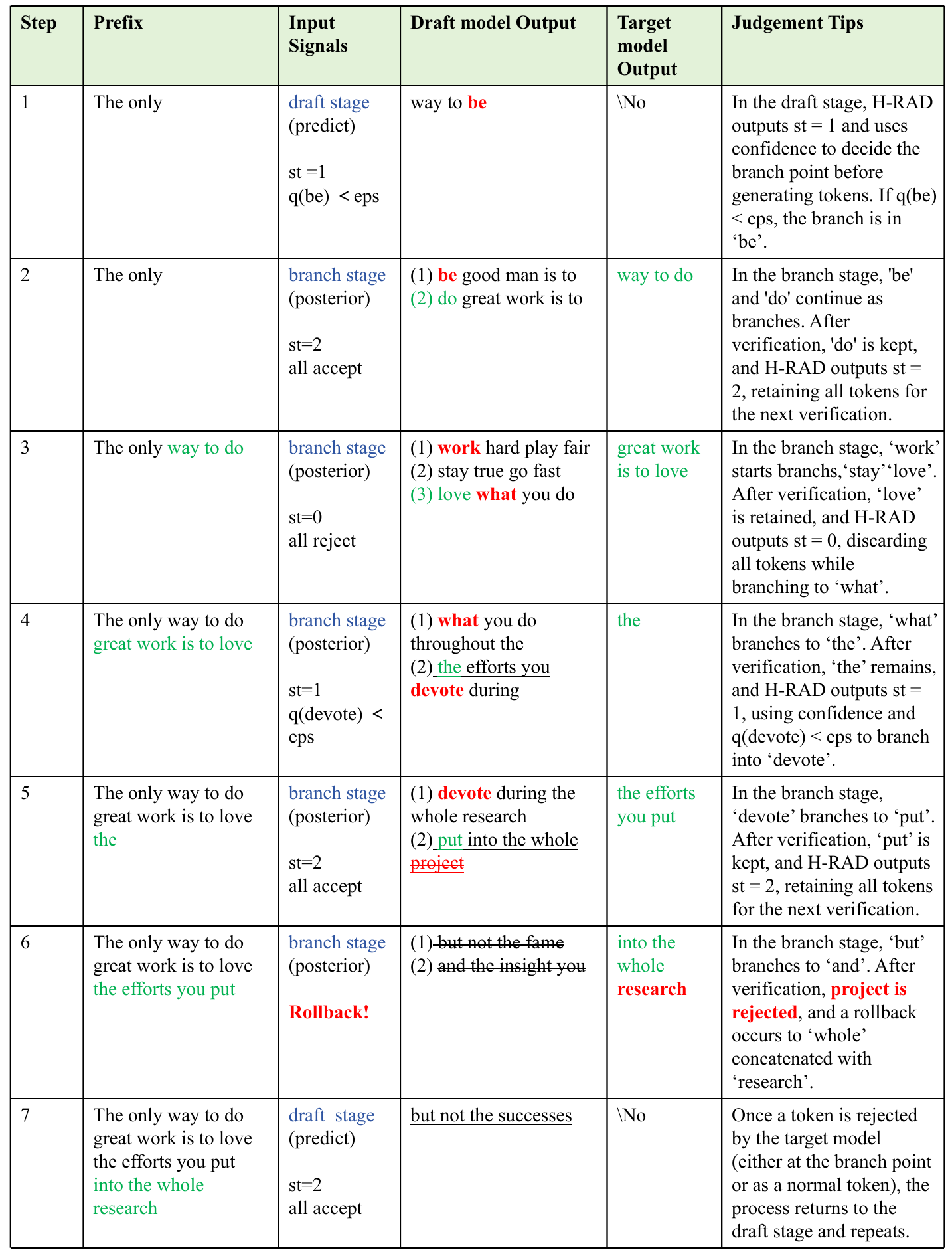}
\vspace*{-0.2in}
\caption{A step-by-step profiling example of SpecBranch with an input prefix `the only'. }
\vspace*{-0.1in}
\label{fig:appendix_example}
\end{figure*}

We have illustrated the main process of SpecBranch in Fig.~\ref{fig:framework}. Here, we provide a more detailed step-by-step profiling example of SpecBranch with an input prefix `the only' in Fig.~\ref{fig:appendix_example}.

The \textcolor{blue}{blue text} represents the stage and the \textcolor{red}{red text} indicates branch points or rejected tokens. \underline{Underlined text} signifies the tokens retained after each round of posterior drafting in the branch stage, which are sent to the target model for verification. The dashed lines represent the tokens discarded during rollback. We give some explanations about the whole process step by step.
\begin{enumerate}[leftmargin=*,align=left]

\item[1)]At step 1, we extract features from the prefix `The only' and input it to H-RAD (assuming the preceding context is already established), which outputs the generation strategy $s_t = 1$. This means we use the implicit confidence $q(x)$ to determine when to begin branching. Then, `The only' is input to the draft model, and the target model does not operate during this stage. The draft model generates `way to' until $q(\text{be}) < \epsilon$, at which point we branch at `be' and send `way to' to the target model for parallel verification, transitioning from the draft stage to the branch stage.

\item[2)]At step 2, `be' and `do' are generated in parallel as two separate branches, while `way to' is verified by the target model. Once both drafting and verification are completed, the target model outputs `way to do,' indicating that the second branch `do' is selected and the `be' branch is discarded. This allows SpecBranch to successfully avoid the rollback of `be' through H-RAD prediction and branch resampling. Then, H-RAD reuses the feature output $s_t = 2$, indicating that all tokens have a high acceptance probability, and `great work is to' is sent for verification in the next round.

\item[3)]At step 3, since $s_t = 2$, `word' is the first token of this round and is considered as an unconfident token, prompting the generation of new branches `stay' and `love'. After the target model verifies and accepts the sequence `great work is to,' it also outputs `love' to match the branch. By generating a new branch, the rejection of `work' is avoided. Then, H-RAD reuses the feature output $s_t = 0$, indicating that all tokens have a low acceptance probability, and only `love' is sent to the target model to match the branch point `what'.
   
\item[4)]At step 4, `what' branches to `the' and generates tokens in parallel. Then, the target model outputs `the', meaning `what' is rejected and the `the' branch is retained, thus avoiding the rollback of `what'. H-RAD and branch resampling combine to improve token rollback handling, while PEARL suffers from token rollback. Then, H-RAD outputs $s_t = 1$, meaning we need to use confidence to determine the branch point `devote', and send `the efforts you' to the target model for the next round of verification.

\item[5)]At step 5, similarly, SpecBranch combats the unconfident token `devote' through implicit confidence early stopping and branch resampling, improving parallel efficiency. `the efforts you put' is accepted, and H-RAD outputs $s_t = 2$ to retain all tokens generated by the `put' branch.

\item[6)]At step 6, we provide a rollback case. In the new round, the first token `but' branches to `and', drafting tokens. However, the target model rejects the previous round's token `project' and resamples a new token `research'. This means that all the tokens generated in parallel need to be discarded, and we return to `whole' and concatenate with `research'. Meanwhile, if none of the branch points match the output of the target model, this is also treated as a rollback.

\item[7)]At step 7, once a token is rejected by the target model (either at the branch point or as a normal token), the process returns to the draft stage and repeats the above process of draft branch in parallel. For those ``doomed tokens'' that cannot be avoided, they have to be rolled back to the draft stage. This completes the feedback loop.

\end{enumerate}

\begin{figure*}[h]
\centering
\includegraphics[width=0.8\linewidth]{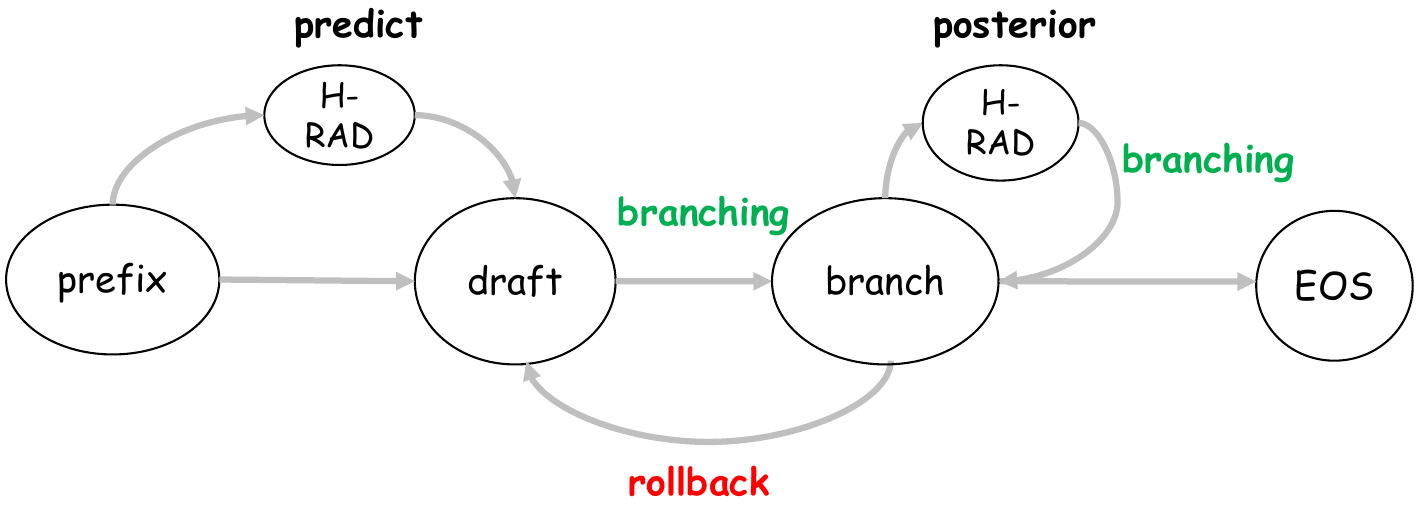}
\caption{Illustration of the SpecBranch state transitions. }
\vspace*{-0.1in}
\label{fig:appendix_logical}
\end{figure*}

Meanwhile, to better illustrate the SpecBranch workflow, we present a simplified stage-transition logical loop in Fig.~\ref{fig:appendix_logical}. Given a prefix, if target-model features are available, H-RAD reuses these features to predict the draft length during the draft stage. The transition from the draft stage to the branch stage occurs when a branch point is selected, which can be one of three scenarios: (1) $s_t = 0$ (all-reject), indicating that the branch point is the first token of this round; (2) $s_t = 1$, where confidence is used to decide the branch point; and (3) $s_t = 2$ (all-accept), indicating that the branch point is the first token of the next round. Each scenario triggers a distinct branching to ensure correct overlap between drafting and verification.

Similarly, in the branch stage, H-RAD uses a posterior approach to select the retained tokens, leading to three branching scenarios for parallel execution. Once a token is rejected by the target model (either at the branch point or as a regular token), the branch stage reverts to the draft stage and repeats the draft-branch process in parallel, thus maintaining the logical loop.

\section{Branch Speculative Sampling} \label{sec:appendix_eval}
As discussed in Section~\ref{sec:branch_resampling}, we verify the branch point $x_b$ using $p(x_b)$ derived from $x_{b-1}$, and apply branch speculative sampling algorithms to sample or adjust the distribution of candidate branches. The branch point verification is formalized as:
\begin{equation}
\mathcal{V} = \text{Match}\bigl( \underbrace{\bigl( q(x_b^1) \ldots, q(x_b^k) \bigr)}_{\text{Draft probability}} , \underbrace{\bigl( p(x_b^1), \ldots, p(x_b^k) \bigr)}_{\text{Target probability}} \bigr) \big\}.
\end{equation}

Single-round speculative sampling relies on a chain-structured draft, whereas SpecBranch adopts a branch structure. Inspired by~\citep{li2024eagle,miao2024specinfer}, SpecBranch introduces Branch Speculative Sampling. Specifically, given the target model's distribution $p(x_b)$ for the branch point, we perform speculative sampling token-by-token on the top-$k$ branch point candidates. For each candidate $x_b^i$, we evaluate it against the criterion $r_b < \frac{p(x_b^i)}{q(x_b^i)}$. Once any branch point token $x_b^i$ is verified and accepted, the corresponding branch is selected and retained. In non-acceptance scenarios, branch speculative sampling recursively invokes single-round speculative sampling, instead of retaining the original naive sampling. The pseudocode corresponding to Branch Speculative Sampling is detailed in Algorithm~\ref {alg:bss}.

\begin{algorithm}
   \caption{Branch Speculative Sampling}
   \label{alg:bss}
\begin{algorithmic}
   \STATE {\bfseries Input:}  Branch points target model distribution $p(x_b)$, top-$k$ branch points $x_b^i$ and distributions $q(x_b^i)$ for each $i$ from 1 to $k$, where $x_b^i$ is sampled from $q(x_b^i)$,
   \STATE {\bfseries Output:} a sample $x_b \sim p(x_b)$;
   \STATE  $i \leftarrow 1$
   \FOR {$i \leq k$}
   \STATE $r_b \leftarrow U(0,1)$ 
   \IF{$r_b<{p(x_b^i)}/{q(x_b^i)}$}
   \STATE {\bfseries Return} $x_b^i$
   \ENDIF
   \STATE $p(x_b) \leftarrow \texttt{norm}(\max(0,p(x_b)-q(x_b^i)))$ 
   \STATE $i \leftarrow i+1$
   \ENDFOR
\STATE Sample a new token $x_b \sim p(x_b)$
\STATE {\bfseries Return} $x_b$
\end{algorithmic}
\end{algorithm}

Unlike tree-based methods that spawn branches at every token (expensive KV-Cache growth)~\citep{miao2024specinfer}, SpecBranch limits branching to uncertainty points identified by H-RAD and avoids the need for complex tree attention mask verification. This means SpecBranch only applies branch speculative sampling at branch points, while employing the typical speculative sampling for subsequent draft tokens. This ensures an \textbf{identical sampling distribution of the target model}, enabling lossless acceleration while also significantly reducing deployment complexity and computational overhead.

Our main experiments are conducted \textbf{under greedy decoding conditions} that are consistent across all baselines i.e., target model temperature $=0$. To further verify the lossless nature of our method, we perform additional experiments on GSM8K across different temperatures to evaluate accuracy. As shown in the Table~\ref{tab:lossless}, SpecBranch achieves identical accuracy with the vanilla autoregressive decoding across various model pairs and temperatures, which further justifies that Branch Speculative Sampling guarantees the original output distribution of the target LLM.

\begin{table}[h]
\centering
\vspace{-2pt}
\setlength{\tabcolsep}{4pt}
\renewcommand{\arraystretch}{0.7} 
\tiny 
\begin{adjustbox}{width=\textwidth}
\begin{tabular}{>{\centering\arraybackslash}p{1.2cm} p{1.2cm} cccccc c c}
\toprule
\multirow{2}{*}{\textbf{Models}} & \multirow{2}{*}{\textbf{Methods}} & \multicolumn{2}{c}{\textbf{Temperature=0}} & \multicolumn{2}{c}{\textbf{Temperature=0.5}} & \multicolumn{2}{c}{\textbf{Temperature=1}}  \\
\cmidrule(lr){3-4} \cmidrule(lr){5-6} \cmidrule(lr){7-8}
& & Acc. & Speedup & Acc. & Speedup & Acc. & Speedup\\

\midrule
\multirow{2}{*}{\centering \begin{tabular}{@{}c@{}} Vicuna\end{tabular}}
                  & Vanilla          & 0.29 & 1.00$\times$ & 0.25 & 1.00$\times$ & 0.21 & 2.11$\times$ \\
  &\textbf{SpecBranch} & \textbf{0.29} & \textbf{1.95}$\times$ & \textbf{0.25} & \textbf{1.83}$\times$ & \textbf{0.21} & \textbf{1.75}$\times$   \\

\midrule
\multirow{2}{*}{\centering \begin{tabular}{@{}c@{}} LLaMA-3.1\end{tabular}}
                  & Vanilla          & 0.93 & 1.00$\times$ & 0.92 & 1.00$\times$ & 0.90 & 1.00$\times$ &\\
  & \textbf{SpecBranch} & \textbf{0.93} & \textbf{3.67}$\times$ & \textbf{0.92} & \textbf{3.58}$\times$ & \textbf{0.90} & \textbf{3.54}$\times$ \\
\bottomrule
\end{tabular}
\end{adjustbox}
\caption{{SpecBranch achieves lossless acceleration across various model pairs and temperatures} on GSM8K~\citep{Cobbe_Kosaraju_Bavarian_Hilton_Nakano_Hesse_Schulman_2021}. } 
\label{tab:lossless}
\end{table}

\section{Evaluation Details} \label{sec:appendix_eval}
For reproducibility, we discuss the experimental setup (Section~\ref{sec:Experiments}) in detail and the source code of this project will be made available at a later time.

\subsection{Data Configurations}
\label{sec:appendix_eval1}

In our experiments, we evaluate SpecBranch using the following dataset settings. The tasks include code generation, multilingual arithmetic reasoning, summarization, and Spec-Bench (a widely-adopted benchmark consisting of six sub-tasks)~\cite{xia2024unlocking}. For code generation, we use HumanEval~\cite{chen2021codex}, a widely recognized benchmark comprising 164 problems. For arithmetic reasoning and multilingual inference, we use GSM8K~\cite{Cobbe_Kosaraju_Bavarian_Hilton_Nakano_Hesse_Schulman_2021}, presenting their results side by side. Specifically, we sample the first 100 examples from GSM8K. For summarization, we use CNN/DM~\cite{Nallapati_Zhou_dos}, and also sample the first 100 examples. The maximum generation length for these tasks is set to 512 tokens. 

For Spec-Bench, we define distinct templates for each subtask. The template for Vicuna follows the official format, while for LLaMA and DeepSeek, the templates are as follows:
\begin{enumerate}[leftmargin=*,align=left]
\item[\ding{71}]\textbf{MT-Bench}: ``A conversation between a curious user and an AI assistant, where the assistant provides helpful, detailed, and polite answers to the user's questions."

\item[\ding{71}]\textbf{QA}: ``A conversation between a curious user and an AI assistant, where the assistant gives helpful, detailed, and polite answers to the user's questions.''

\item[\ding{71}]\textbf{Summarization}: ``Summarize: {{QUESTION}} TL;DR.''

\item[\ding{71}]\textbf{Translation}: ``Translate German to English. German: {{QUESTION}} English.''

\item[\ding{71}]\textbf{Math}: ``Let's think step by step.''

\item[\ding{71}]\textbf{RAG}: ``A conversation between a curious user and an AI assistant, where the assistant gives helpful, detailed, and polite answers to the user's questions.''
\end{enumerate}

Specifically, due to the unique characteristics of LLaMA-3.1, we design a specialized template as follows: ``You are a helpful, respectful, and honest assistant. Always answer as helpfully as possible, while being safe. Your answers should not include any harmful, unethical, racist, sexist, toxic, dangerous, or illegal content. Please ensure that your responses are socially unbiased and positive in nature. If a question does not make sense or is not factually coherent, explain why instead of providing an incorrect answer. If you don't know the answer, please do not share false information.''

\subsection{Model Configurations}
\label{sec:appendix_eval2}

To validate performance, we select state-of-the-art open-source model pairs such as the LLaMA series (JackFram
/llama-68m, huggyllama/llama-7b), Vicuna (double7/vicuna-68m, lmsys/vicuna-13b-v1.3), Deepseek-Coder (deepseek-ai/deepseek-coder-1.3b-instruct, deepseek-ai/deepseek-coder-33b-instruct) and LLaMA-3.1 (meta-llama/Llama-3.1-8B-Instruct, meta-llama/Llama-3.1-70B-Instruct) for each task. All model weights are loaded in bfloat16 format for optimized GPU inference without quantization. As a draft model training-free method, SpecBranch does not modify any draft model parameters during evaluation. We summarize the model configuration in Table~\ref{tab:model_conf}.

\begin{table}[h]
\centering
\small  
\resizebox{0.8\columnwidth}{!}{  
\begin{tabular}{ccccc}
\toprule
    \textbf{Models} & \textbf{Layers} & \textbf{dim} & \textbf{FFN dim} & \textbf{Vocabulary size} \\
\midrule
LLaMA 68M & 2  & 768  & 3072  & 32000   \\
LLaMA 7B & 32  & 4096  & 11008  & 32000  \\
Vicuna 68M & 2  & 768  & 3072  & 32000    \\
Vicuna 13B & 40  & 5120  & 13824  & 32000  \\
Deepseek 1.3B & 24  & 2048  & 5504  & 32256   \\
Deepseek 33B & 62 & 7168  & 19200  & 32256   \\
LLaMA-3.1 8B & 32 & 4096 & 14336  & 128256   \\
LLaMA-3.1 70B & 80 & 8192  & 28672  & 128256 \\
\bottomrule
\end{tabular}
}
\vspace*{0.1in}
\caption{Model configurations.}
\label{tab:model_conf}
\vspace*{-0.1in}
\end{table}

\subsection{Evaluation Details}
\label{sec:appendix_eval3}
We report widely used metrics for speculative decoding (SD): Mean Accepted Length $M$, Wall-Time Speedup Ratio, and Speed (tokens/sec). Additionally, we introduce a new metric, Rollback Rate (RB), defined as \( RB = \frac{\# \text{Rollback tokens}}{\# \text{Total tokens}}, \) which quantifies computational waste due to invalid drafts. In SpecBranch, $M$ represents the continuously accepted length~\citep{liu2024parallel}, which is not the fixed length accepted in a single round of $\gamma$, but rather the higher accepted length achieved through multiple rounds of parallel generation, surpassing the performance of Vanilla SD. $RB$ specifically refers to the number of rollbacks during the draft model’s forward times, excluding additional token loss due to branch and tree structures (since the impact of draft parallelism on acceleration is negligible).

All experiments, including the main results and ablation studies, are conducted on NVIDIA A100-PCIE-40G GPUs. Models with fewer than $8$B parameters are run on a single device, $33$B models on two devices, and $70$B models on four devices. For inference, we set the batch size to $1$ to match standard speculative decoding settings. 

\textbf{Temperature Sampling}\quad Since Top-$k$ sampling is adopted, we set the draft model temperature to $1$ and apply a greedy sampling method with the target model temperature set to $0$. For other baselines, we set the draft and target model temperature to $0$ as greedy sampling. 

\textbf{Resource Consumption}\quad In the resource consumption experiments, we use the \texttt{NVIDIA Data Center GPU Manager (DCGM)} to monitor real-time GPU memory usage and power. Energy consumption is calculated by multiplying the average power by the total inference time over the entire benchmark.

 All baselines, such as PEARL and SpS, are reproduced from their original papers and official codebases, using the optimal configurations reported by the authors. \textbf{Experiments are conducted under identical conditions to avoid introducing bias.} Importantly, for LLaMA-3.1 model pairs, the implementation of Lookahead requires transformers $=4.36.2$, which conflicts with LLaMA 3.1’s dependency on transformers $\geq 4.43.0$. A similar situation is also found in~\citep{liu2024parallel}.

\subsection{Training Details}
\label{sec:appendix_eval4}
\label{sec:appendix_eval4}
The training data for H-RAD pairs the feature vector $z_t$ from Eq.\eqref{eq:z_t} with the corresponding three-class labels $s_t$. We implement a lightweight three-layer MLP with ReLU activation and dropout (rate$=0.4$). The model architecture consists of two hidden layers (256 and 64 units) followed by a classification layer. Training is performed \emph{offline} for 20 epochs with 32 batch size using the AdamW optimizer~\cite{loshchilov2017decoupled} (learning rate$=5\times10^{-5}$, weight decay$=1\times10^{-4}$).
To address class imbalance issues, we employ SMOTE (Synthetic Minority Over-sampling Technique) data augmentation~\cite{chawla2002smote}, specifically targeting underrepresented classes in H-RAD. The augmentation process involves standardizing features, applying SMOTE with $k=5$ nearest neighbors, and then inverse transforming the synthetic samples. Additionally, we utilize label smoothing (smoothing$=0.1$) in the loss function to prevent overconfident predictions and improve generalization.
The training process incorporates several optimization strategies:
\begin{enumerate}[leftmargin=*,align=left]
    \item[\ding{71}] Learning rate scheduling with ReduceLROnPlateau (factor$=0.5$, patience$=2$).
    \item[\ding{71}] Early stopping with $5$ epochs patience.
    \item[\ding{71}] Gradient clipping with max norm of $1.0$.
    \item[\ding{71}] Weighted random sampling to balance class distributions.
\end{enumerate}
The training converges within $5$ minutes on a single NVIDIA A100 GPU and avoids the need for costly online fine-tuning while maintaining adaptability to diverse domains. The model achieves balanced performance across the three classes, as validated through confusion matrix analysis and t-SNE visualization of the learned feature space.

\section{More Experimental Results} \label{sec:appendix_results}

\subsection{The Optimal Draft Length }
\label{sec:appendix_length}

\begin{figure*}[h]
\vspace*{-0.1in}
\centering
\includegraphics[width=0.6\linewidth]{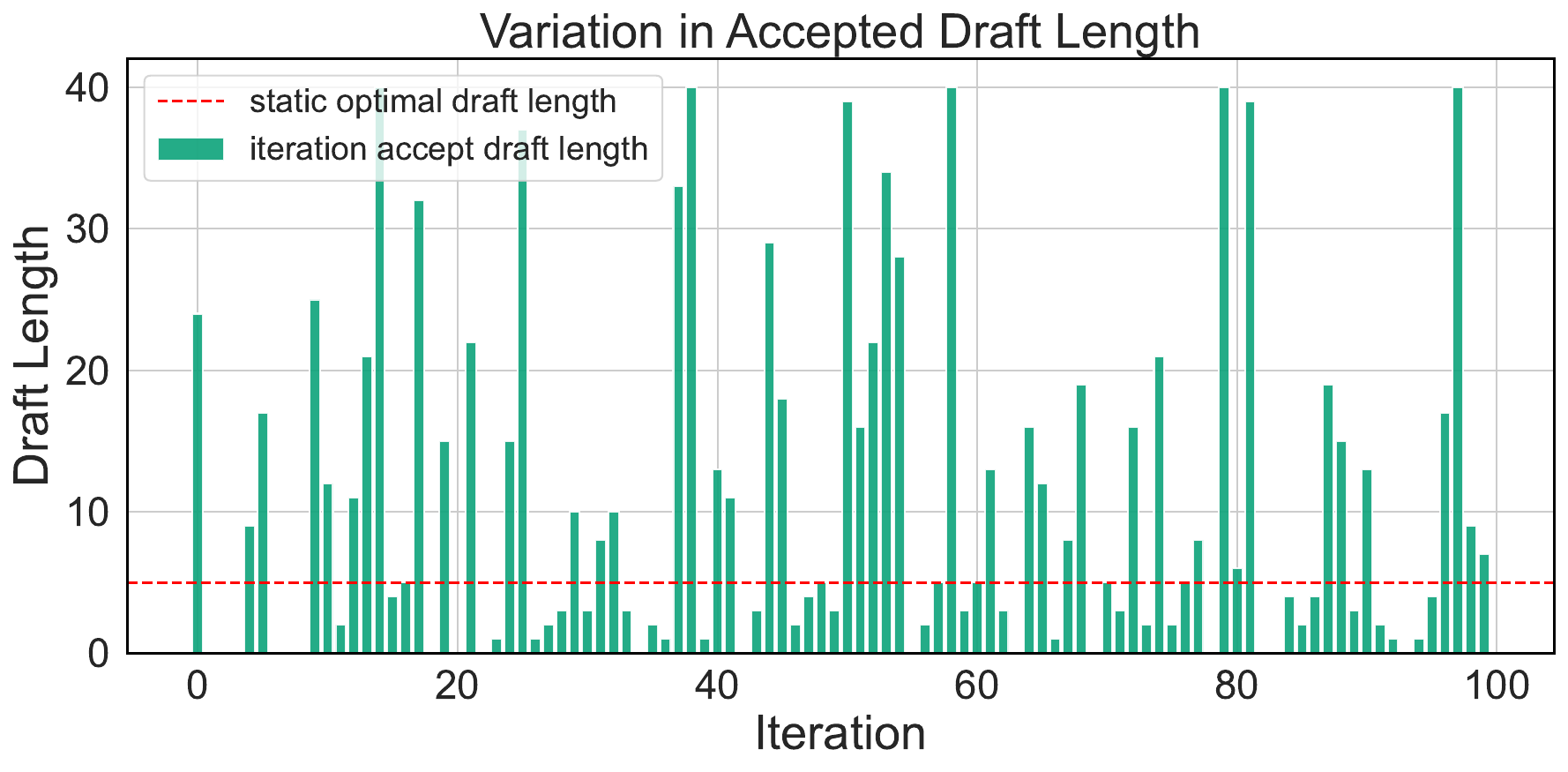}
\vspace*{-0.05in}
\caption{Variation of the optimal draft length over different iterations.}
\label{fig:appendix_length_change}
\end{figure*}

In Section~\ref{sec:analysis1}, we have discussed the theoretical solution under rollback. However, this theoretical analysis only reflects the statistical properties rather than runtime dynamics. As shown in Fig.~\ref{fig:appendix_length_change}, the actual optimal accepted draft length is context-dependent and varies significantly across different iterations and the same phenomenon is observed in~\cite{zhang2024adaeagle,liu2024parallel}. Hence, fixed draft length would result in substantial rollback and token waste. This highlights the need for adaptive control of $\gamma$ rather than relying on static configurations.

\subsection{More Evaluation Results on the Spec-Bench Benchmark}
\label{sec:appendix_result1}

As illustrated in Section~\ref{sec:Experiments}, we provide more evaluation results of SpecBranch in Table~\ref{tab:more_spec} with both LlaMA 68m\&7B and Deepseek 1.3\&33B on Spec-Bench. Notably, Llama 3.1 is a more recent LLM version that requires the transformer version to be greater than $4.43.0$. Due to this incompatibility, we cannot reproduce the results of baseline Lookahead Decoding as described before.

\begin{table*}[h]
  \centering
  \vspace{-2pt}
  \resizebox{\textwidth}{!}{
    \begin{tabular}{c@{\hspace{0.2cm}}l@{\hspace{0.2cm}}c@{\hspace{0.2cm}}c@{\hspace{0.2cm}}c@{\hspace{0.2cm}}c@{\hspace{0.2cm}}c@{\hspace{0.2cm}}c@{\hspace{0.2cm}}c@{\hspace{0.2cm}}c@{\hspace{0.2cm}}c@{\hspace{0.2cm}}c@{\hspace{0.2cm}}c@{\hspace{0.2cm}}c@{\hspace{0.2cm}}c}
    \toprule
    \multirow{2}[4]{*}{\textbf{Models}} & \multirow{2}[4]{*}{\textbf{Methods}} &   \multicolumn{2}{c}{\textbf{MT Bench}}   &   \multicolumn{2}{c}{\textbf{QA}} &  \multicolumn{2}{c}{\textbf{Sum}}  &  \multicolumn{2}{c}{\textbf{Math}} & \multicolumn{2}{c}{\textbf{RAG}} & \multicolumn{2}{c}{\textbf{Trans}} &   \multirow{2}[4]{*}{\textbf{Avg.}} \\
\cmidrule(lr){3-4}  \cmidrule(lr){5-6} \cmidrule(lr){7-8} \cmidrule(lr){9-10} \cmidrule(lr){11-12} \cmidrule(lr){13-14}         &         & M & Speedup & M & Speedup  & M & Speedup & M & Speedup & M & Speedup & M & Speedup &  \\
    \midrule
    \multirow{5}{*}{\large LLaMA} & SpS  &  2.78   &    1.77$\times$      &   4.43    &   2.57$\times$ & 2.62 &  1.51$\times$  & 5.02 & 2.50$\times$ &  2.39 &  1.28$\times$  & 5.28&  2.90$\times$ & 2.09$\times$ \\
          & AdaEDL  &  2.74  &  {1.86$\times$}  &   4.25   &  2.59$\times$ & 2.56 &  1.54$\times$  & 4.83 & 2.43$\times$ &  2.27 &  1.39$\times$  & 4.93 &  2.96$\times$ & 2.13$\times$ \\ 
          & Lookahead&  1.58   &  {1.31$\times$}    &   1.62  &   1.34$\times$ & 1.59 &  1.32$\times$  & 1.73 & 1.51$\times$ &  1.31 &  1.15$\times$  & 1.36&  1.23$\times$ & 1.31$\times$ \\ 
          & PEARL  &  2.83  &  {2.14$\times$}  &   5.98   & 3.15$\times$ & 2.79 &  \textbf{1.85}$\times$  & 7.58 & 3.37$\times$ &  2.48 &  1.43$\times$  & 7.65 &  4.03$\times$ & 2.66$\times$ \\
        & \cellcolor{mygreen} \textbf{SpecBranch} & \cellcolor{mygreen} \textbf{3.01} & \cellcolor{mygreen} \textbf{\cellcolor{mygreen} \textbf{2.34}\(\times\)} & \cellcolor{mygreen} \textbf{6.57} & \cellcolor{mygreen} \textbf{\cellcolor{mygreen} \textbf{3.53}\(\times\)} & \cellcolor{mygreen} \textbf{2.87} & \cellcolor{mygreen}1.76\(\times\) & \cellcolor{mygreen} \textbf{8.57} & \cellcolor{mygreen} \textbf{\cellcolor{mygreen} \textbf{3.77}\(\times\)} & \cellcolor{mygreen} \textbf{3.06} & \cellcolor{mygreen} \textbf{\cellcolor{mygreen} \textbf{1.64}\(\times\)} & \cellcolor{mygreen} \textbf{8.57} & \cellcolor{mygreen} \textbf{\cellcolor{mygreen} \textbf{4.32}\(\times\)} & \cellcolor{mygreen} \textbf{\cellcolor{mygreen} \textbf{2.89}\(\times\)} \\
    \midrule
        \multirow{5}{*}{\large Vicuna} & SpS  &  2.63   &    1.74$\times$      &   2.47    &   1.64$\times$ & 2.58 &  1.70$\times$  & 2.37 & 1.55$\times$ &  2.47 &  1.56$\times$  & 2.57&  1.65$\times$ & 1.64$\times$ \\
          & AdaEDL  &  2.50  &  {1.80$\times$}  &   2.43   &  1.67$\times$ & 2.64 &  1.72$\times$  & 2.31 & 1.62$\times$ &  2.21 &  1.57$\times$  & 2.45 &  1.75$\times$ & 1.69$\times$ \\ 
          & Lookahead&  1.56   &  1.31$\times$    &   1.41   &   1.23$\times$ & 1.49 &  1.25$\times$  & 1.71 & 1.46$\times$ &  1.38 &  1.15$\times$  & 1.32&  1.10$\times$ & 1.25$\times$ \\ 
          & PEARL  &  2.62  &  {1.78$\times$}  &   2.45   &  1.64$\times$ & \textbf{2.78} &  \textbf{1.83}$\times$  & 2.63 & 1.67$\times$ &  2.61 &  1.66$\times$  & 2.89 &  2.05$\times$ & 1.77$\times$ \\
          & \cellcolor{mygreen} \textbf{SpecBranch} &  \cellcolor{mygreen} \textbf{3.11}  & \cellcolor{mygreen} \textbf{2.09}$\times$  &   \cellcolor{mygreen} \textbf{2.67}  &   {\cellcolor{mygreen} \textbf{1.83}$\times$}& \cellcolor{mygreen}2.72 &  \cellcolor{mygreen}1.78$\times$  & \cellcolor{mygreen} \textbf{2.86} & \cellcolor{mygreen} \textbf{1.89$\times$} &  \cellcolor{mygreen} \textbf{2.83} &  \cellcolor{mygreen} \textbf{1.86$\times$}  & \cellcolor{mygreen} \textbf{3.32} &  \cellcolor{mygreen} \textbf{2.30$\times$} & \cellcolor{mygreen} \textbf{1.96$\times$} \\
    \midrule
            \multirow{5}{*}{\large Deepseek} & SpS  &  3.99   &    2.03$\times$      &   4.02    &   2.12$\times$ & 4.08 &  2.02$\times$  & 3.93 & 2.06$\times$ &  3.95 &  1.94$\times$  & 4.37&  2.21$\times$ & 2.06$\times$ \\
          & AdaEDL  &  3.54  &  {2.26$\times$}  &   3.81   &  2.33$\times$ & 3.74 &  2.03$\times$  & 3.81 & 2.28$\times$ &  3.69 &  2.03$\times$  & 4.12 &  2.38$\times$ & 2.22$\times$ \\ 
          & Lookahead&  1.78   & 1.51$\times$     &   1.81   &   1.64$\times$ & 1.59 &  1.42$\times$  & 1.75 & 1.53$\times$ &  1.53 &  1.32$\times$  & 1.74&  1.50$\times$ & 1.49$\times$ \\ 
          & PEARL  &  6.60  &  {2.77$\times$}  &   5.15   &  \textbf{2.91$\times$} & 5.81 &  2.68$\times$  & 7.14 & 2.79$\times$ &  5.77 &  2.65$\times$  & 6.20 &  3.09$\times$ & 2.81$\times$ \\
         & \cellcolor{mygreen} \textbf{SpecBranch} & \cellcolor{mygreen} \textbf{8.31}  & \cellcolor{mygreen} \textbf{3.02$\times$}  &   \cellcolor{mygreen} \textbf{5.71}  &  \cellcolor{mygreen} {2.87$\times$}& \cellcolor{mygreen} \textbf{6.67} &  \cellcolor{mygreen} \textbf{\cellcolor{mygreen} \textbf{2.95}\(\times\)}  & \cellcolor{mygreen} \textbf{8.92} & \cellcolor{mygreen} \textbf{\cellcolor{mygreen} \textbf{3.21}\(\times\)} & \cellcolor{mygreen} \textbf{6.82} &  \cellcolor{mygreen} \textbf{\cellcolor{mygreen} \textbf{2.76}\(\times\)}  & \cellcolor{mygreen} \textbf{6.86} &  \cellcolor{mygreen} \textbf{\cellcolor{mygreen} \textbf{3.28}\(\times\)} & \cellcolor{mygreen} \textbf{\cellcolor{mygreen} \textbf{3.02}\(\times\)} \\
    \midrule
    \multirow{5}{*}{\large LLaMA-3.1} & SpS &  4.67  &   2.31$\times$    &     4.57       &  2.27$\times$ & 5.09 & 1.98$\times$ & 5.01 & 2.44$\times$& 5.08 & 2.02$\times$ & 5.52 & 2.57$\times$ & 2.27$\times$\\
          & AdaEDL  &   4.31   &  2.43$\times$     &  4.23   & 2.30$\times$ &  4.83 & 2.05$\times$ & 4.94 & 2.46$\times$ & 4.86 & 2.13$\times$ & 5.24 & 2.65$\times$ & 2.34$\times$ \\ 
          & Lookahead &  -  &  - &   -   &  - & - &  - & - & - &  - &  -  & - &  - & - \\ 
          & PEARL &  8.46  &  {2.96$\times$}  &   8.37   &  3.27$\times$ & 9.10&  3.32$\times$  & 12.53 & 3.39$\times$ &  8.35 &  \textbf{3.41}$\times$  & 12.59 &  4.22$\times$ & 3.43$\times$ \\
        & \cellcolor{mygreen} \textbf{SpecBranch}&\cellcolor{mygreen} \textbf{10.85} & \cellcolor{mygreen} \textbf{3.24\(\times\)} &\cellcolor{mygreen} \textbf{10.59} & {\cellcolor{mygreen} \textbf{3.45\(\times\)}} & \cellcolor{mygreen} \textbf{11.40} & \cellcolor{mygreen} \textbf{3.63\(\times\)} & \cellcolor{mygreen} \textbf{15.76} & \cellcolor{mygreen} \textbf{3.78\(\times\)} & \cellcolor{mygreen} \textbf{9.16} &\cellcolor{mygreen} 3.40\(\times\) & \cellcolor{mygreen} \textbf{16.64} & \cellcolor{mygreen} \textbf{4.51\(\times\)} & \cellcolor{mygreen} \textbf{3.67\(\times\)} \\
    \bottomrule
    \end{tabular}%
    }
  \label{tab:more_spec}%
\caption{More evaluation results on Spec-Bench. ``--'' indicate incompatibility: baseline implementations (transformers $=4.36.2$) and LLaMA 3.1’s dependency on $\geq 4.43.0$.}
\vspace{-1.5pt}
\end{table*}

We draw several interesting findings from Table~\ref{tab:more_spec}: 1) SpecBranch demonstrates consistent speedups across six subtasks, regardless of the model alignment quality; 2) In particular, the LLaMA 68M \& 7B model combination performs exceptionally well, due to the capacity of the LLaMA model, which excels in the Math, QA, and Translation tasks, with an average accepted length even surpassing DeepSeek. However, there are significant performance variations across other tasks. This suggests that the alignment between the small draft models and the large target model may vary across tasks, leading to highly task-specific effects. These findings also indicate incorporating multiple draft models for speculative sampling in future research. Different draft models can be tailored to specific tasks under the Mixture-of-Experts (MoE) principles to further enhance inference efficiency. 

\begin{figure*}[h]
\vspace*{-0.1in}
\centering
\includegraphics[width=1.00\linewidth]{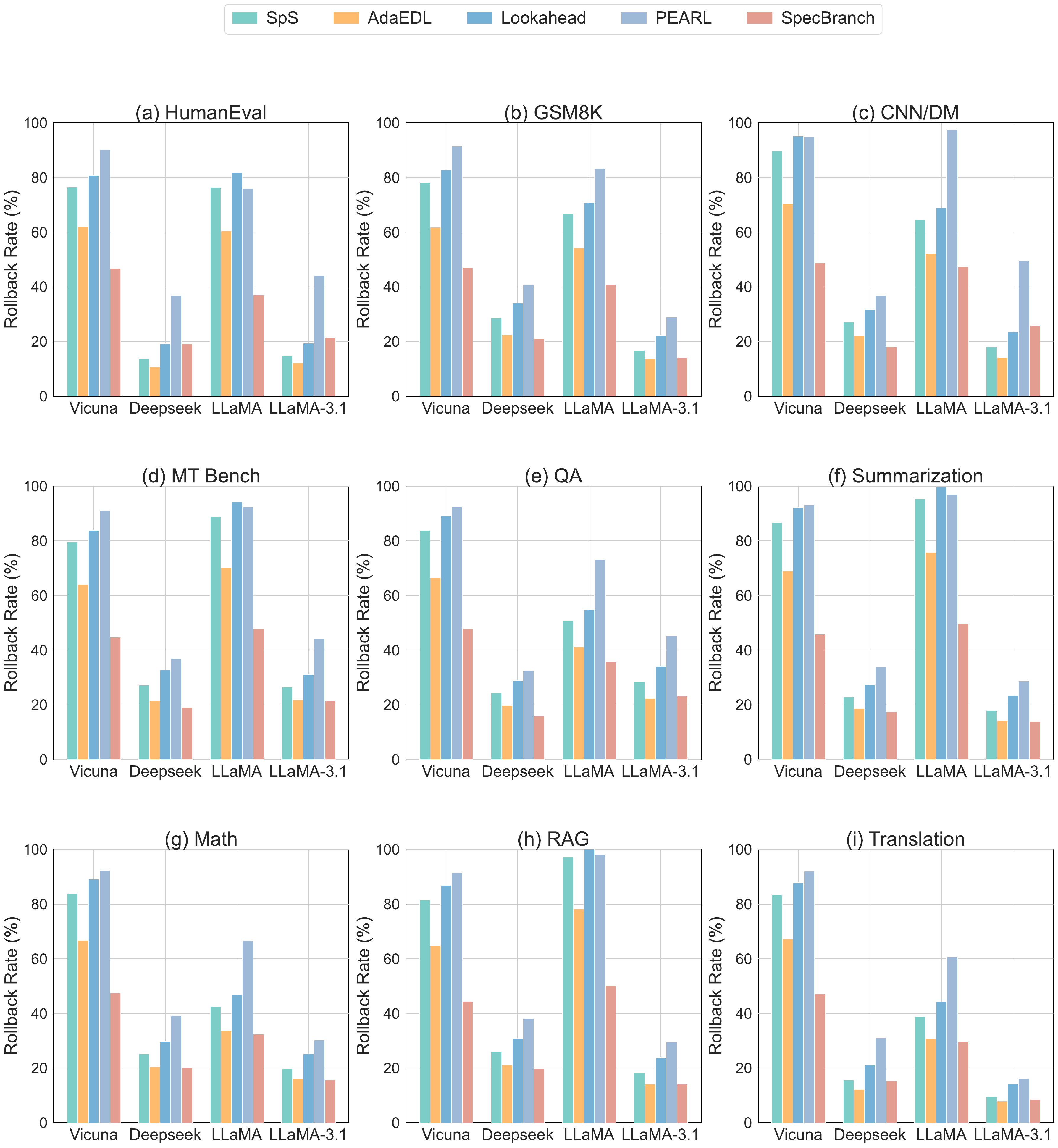}
\vspace*{-0.2in}
\caption{Comparison of Rollback Rates on HumanEval, GSM8K, CNN/DM, Spec-Bench for different model combinations.}
\vspace*{-0.1in}
\label{fig:rollback_grid1}
\vspace*{-0.05in}
\end{figure*}

\subsection{More Evaluation Results of Rollback Ratio}
\label{sec:appendix_result3}
As discussed in Section~\ref{sec:Experiments}, rollback is an important metric for the parallel efficiency. Here, we provide more evaluation results on HumanEval, GSM8K, CNN/DM, and Spec-Bench, as shown in Fig.~\ref{fig:rollback_grid1}. The results demonstrate that SpecBranch achieves significantly lower rollback ratios across various datasets and subtasks compared to other methods. This is consistent with our prior justification that the H-RAD module plays an essential role in mitigating rollback -- the additional evaluations in Fig.~\ref{fig:rollback_grid1} further validate the generalization capacity of H-RAD in different tasks. It is worth mentioning that for poorly aligned model combinations (e.g., Vicuna, LLaMA), SpecBranch reduces the rollback ratio by nearly 50\% compared to PEARL. Even for better-aligned model pairs (e.g., Deepseek, LLaMA-3.1), it achieves about 10\% reduction. These results indicate the potential of SpecBranch in resource-constrained environments that the draft model sizes are typically restrained. The proposed framework can reduce the percentage of rollbacks as well as save computation/energy resources in the long run. 

\begin{figure*}[h]
\vspace*{-0.1in}
\centering
\includegraphics[width=1.00\linewidth]{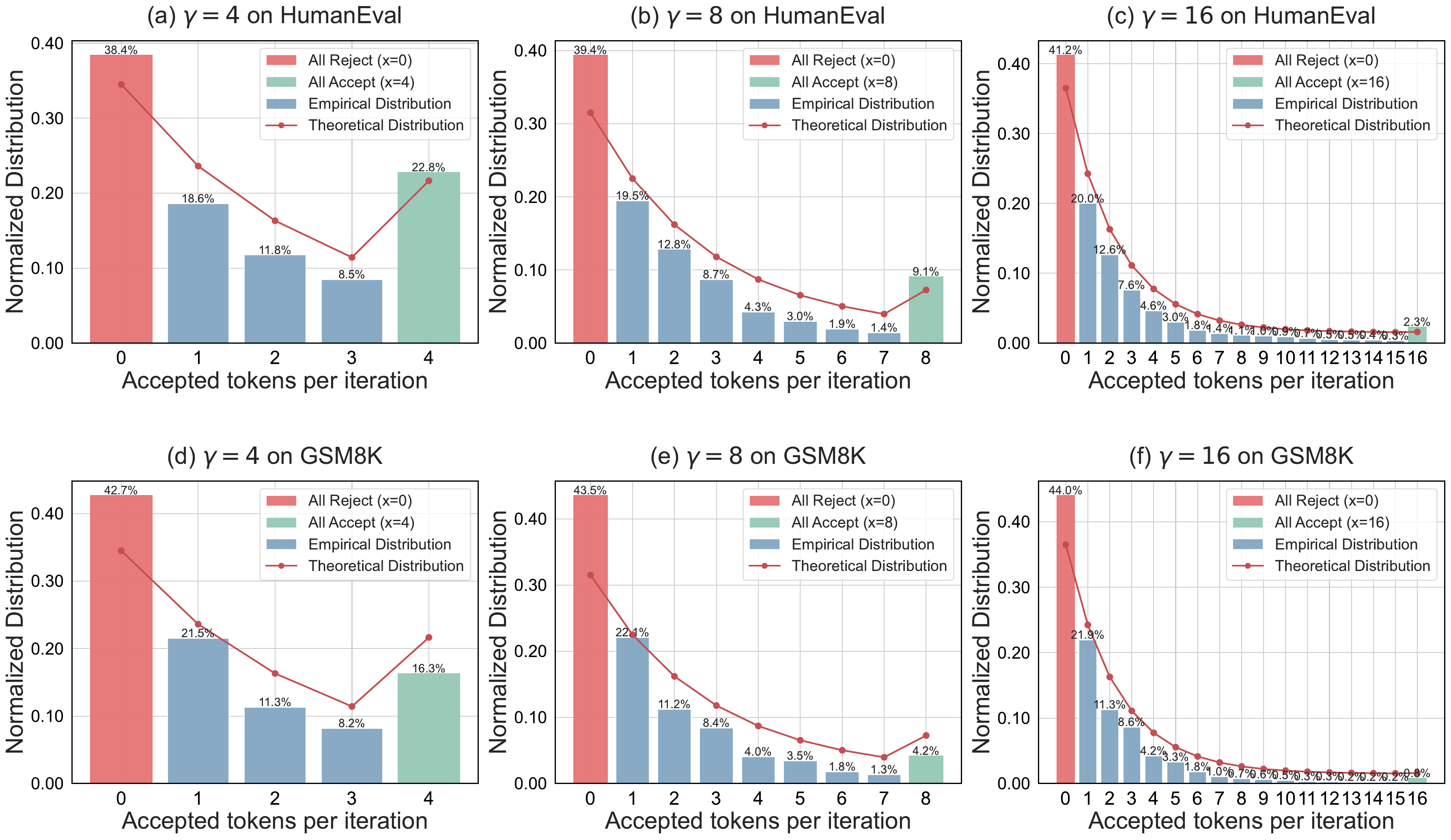}
\vspace*{-0.2in}
\caption{More evaluation results demonstrate that the distribution of accepted tokens generally follows a truncated geometric distribution of different token length $\gamma$ of Vicuna 68M \& 13B.}
\vspace*{-0.05in}
\label{fig:appendix_token_distribution1}
\end{figure*}

\begin{figure*}[h]
\centering
\includegraphics[width=1.00\linewidth]{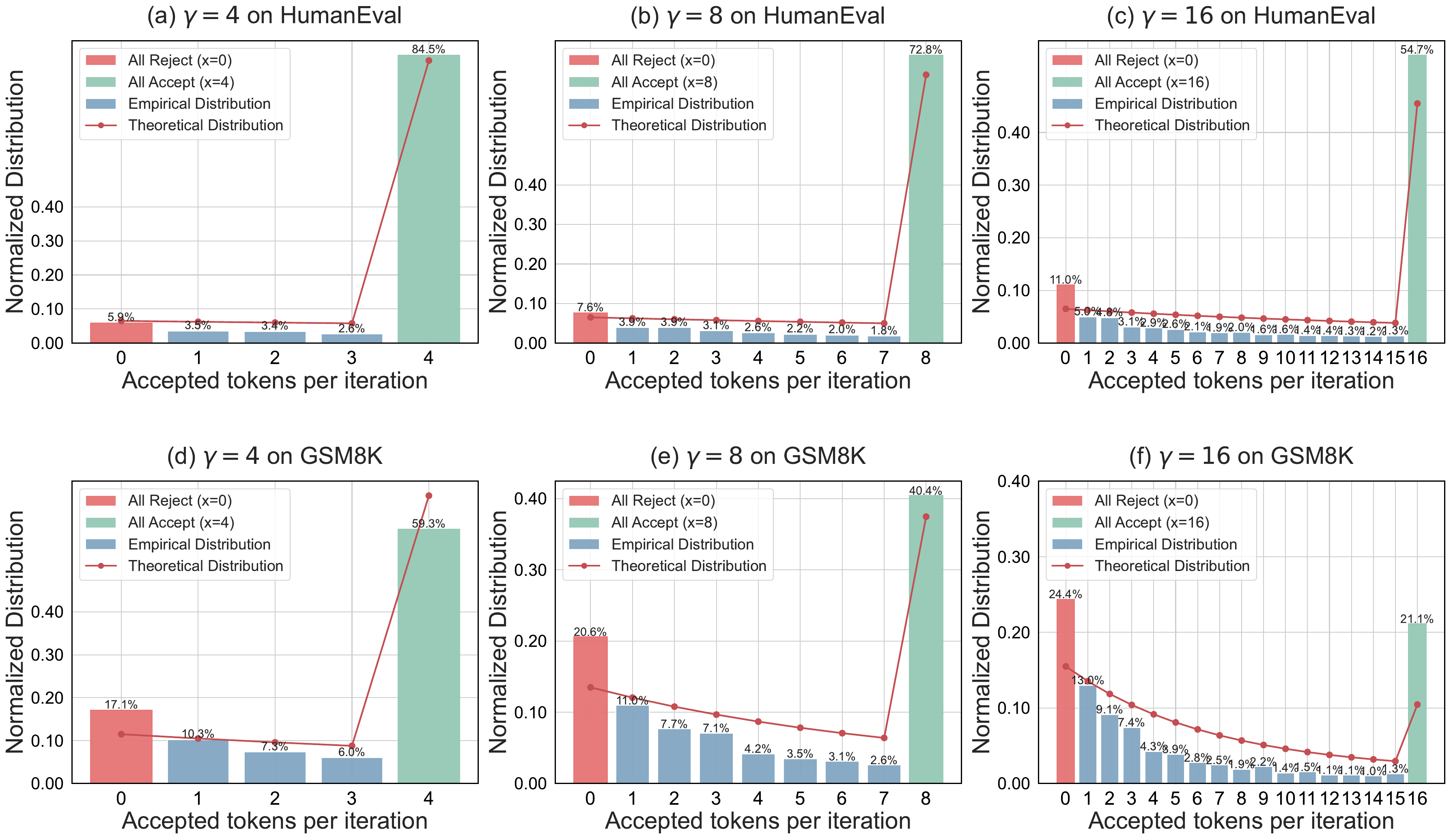}
\vspace*{-0.2in}
\caption{Distribution of accepted tokens generally follows a truncated geometric distribution of different token length $\gamma$ of Deepseek 1.3B \& 33B.}
\vspace*{-0.1in}
\label{fig:appendix_token_distribution2}
\vspace*{-0.13in}
\end{figure*}

\subsection{More Results of Token Distribution}
\label{sec:appendix_result4}
Recall that in Section~\ref{sec:analysis1} and Fig.~\ref{fig:1}(b), we introduce a truncated geometric distribution~\cite{leviathan2023fast}(shown in Fig.~\ref{fig:1}(b),
\begin{equation}
P(X = k) = (1 - \alpha) \cdot \alpha^k \cdot \mathbb{I}(k < \gamma) + \alpha^\gamma \cdot \mathbb{I}(k = \gamma),
\end{equation}
where $\alpha^{\gamma}$ is the probability of \textit{full acceptance} and $1-\alpha^{\gamma}$ is the probability of \textit{rollback}. 
To validate this, we present additional token distribution results for Vicuna 68M \& 13B and Deepseek 1.3\&33B on HumanEval and GSM8K in Figs.~\ref{fig:appendix_token_distribution1} and~\ref{fig:appendix_token_distribution2}. These results show that the token acceptance distribution closely follows the truncated geometric distribution, which is consistent with our prior statements of a bimodal phenomenon extracted from the target model features. This has laid the foundation for the H-RAD module to perform token length predictions with high fidelity.



In particular, Fig.~\ref{fig:appendix_token_distribution1} shows that for poorly aligned models, the acceptance rate $\alpha$ is low, and the truncated geometric distribution peaks at All-Reject. In this scenario, the rollback in SD is exacerbated, particularly in Parallel SD, since it only achieves parallelism under the All-Accept condition. Given the low proportion of All-Accept in this token distribution, the efficiency of parallelism is significantly impacted by rollback. This highlights our motivation to jointly consider rollback with parallelism for traditional SD.

For better-aligned models, Fig.~\ref{fig:appendix_token_distribution2} shows that the acceptance rate $\alpha$ is high and the truncated geometric distribution peaks at All-Accept. In this scenario, the rollback effects in SD are undermined, with the balance between rollback and parallelism leaning towards parallelism. As a result, the parallel framework shows significant acceleration compared to vanilla SD, even approaching the theoretical speedup limit with well-aligned models. However, it is evident that as the draft length increases, the impact of rollback becomes non-negligible. SpecBranch effectively balances the trade-off between these factors.

\subsection{Time and Energy Consumption}
\label{sec:appendix_result5}

We have provided some evaluation results of time and energy consumption in Section~\ref{sec:ablation}. In this section, we conduct more experiments on the time consumption of various model pairs on HumanEval while we further test the energy consumption of various model pairs on HumanEval and GSM8K. 

\textbf{Time consumption} \quad As shown in Table~\ref{tab:app:time_cost}, regardless of model size, the time spent on H-RAD prediction and communication between multiple GPUs is almost negligible compared to the total inference time for a single step. This indicates that the H-RAD module retains its lightweight nature with minimum resource consumption and the inter-GPU communications have low operational overhead as well. On the other hand, the time span for the draft stage and the target model verification stage is nearly equal. This is consistent with our prior results that SpecBranch effectively implements parallelism between these two stages to alleviate the mutual waiting bubbles. 

\begin{table}[h]
\centering
\resizebox{\columnwidth}{!}{  
\begin{tabular}{ccccc}
\toprule
\textbf{Modules} & \textbf{LLaMA 68M\&7B} & \textbf{Vicuna 68M\&13B} & \textbf{Deepseek 1.3\&33B} & \textbf{LLaMA-3.1 8B\&70B}\\
\midrule
\rowcolor{mygreen}\textbf{H-RAD Predict} & \textbf{0.26 ms }& \textbf{0.27 ms} & \textbf{0.31 ms} & \textbf{0.28 ms} \\
Communication & 0.21 ms & 0.31 ms & 0.26 ms & 0.26 ms \\
Draft Stage & 20.8 ms & 30.9 ms & 58.3 ms & 125.1 ms \\
Verification Stage & 21.7 ms & 31.4 ms & 56.1 ms & 128.0 ms \\
\bottomrule
\end{tabular}
}
\caption{Time cost of each module (per step) on the HumanEval dataset.}
\label{tab:app:time_cost}
\vspace*{-0.1in}
\end{table}

\textbf{Energy consumption} \quad
We use the \texttt{NVIDIA Data Center GPU Manager (DCGM)} toolkit to monitor the real-time power consumption and collect relevant traces. Energy consumption is calculated by multiplying the average power by the total inference time over the entire benchmark. The results are provided in Table~\ref{tab:app: energy_human} and Table~\ref{tab:app: energy_gsm9k}. 

For poorly aligned models (LLaMA, Vicuna), SpecBranch with its rollback-aware dynamic draft length through H-RAD, significantly reduces redundant tokens, thereby lowering energy consumption compared to PEARL and SpS. 
Unlike PEARL, which pre-verifies only the first token using the target model in parallel, SpecBranch utilizes H-RAD to predict all tokens during the draft stage. This approach reduces the number of forward passes required by the target model. For LLaMA 68M\&7B on HumanEval, SpecBranch reduces the target model forward passes to \textbf{32,285}, compared to \textbf{44,949} in PEARL, resulting in lower energy consumption for target model inference. For better-aligned models (Deepseek, LLaMA-3.1), where token rollback is significantly reduced, parallel efficiency improves considerably. While energy savings are less pronounced, SpecBranch still outperforms PEARL.

Theoretically, Parallel SD incurs negligible energy consumption compared to Vanilla SD. The speedup achieved through parallelism offsets the additional power consumption caused by extra forward passes in the target model. However, the parallelism and communication overhead still contribute to some energy consumption in real-world deployments. Overall, SpecBranch, with its hybrid rollback-aware draft structure, significantly optimizes energy consumption compared to PEARL. Notably, for poorly aligned models, it outperforms SD in terms of energy efficiency.

\begin{table}[h]
\centering
\resizebox{\columnwidth}{!}{  
\begin{tabular}{ccccc}
\toprule
\textbf{Methods} & \textbf{LLaMA 68M\&7B} & \textbf{Vicuna 68M\&13B} & \textbf{Deepseek 1.3\&33B} & \textbf{LLaMA-3.1 8B\&70B}\\
\midrule
SpS & 217 KJ & 383 KJ & 565 KJ & 1021KJ \\
PEARL & 287 KJ & 445 KJ & 631 KJ & 1231 KJ \\
\rowcolor{mygreen}\textbf{SpecBranch} & \textbf{156 KJ }& \textbf{251 KJ} & \textbf{582 KJ} & \textbf{1092 KJ} \\
\bottomrule
\end{tabular}
}
\caption{Energy cost of SpecBranch and baseline methods on the HumanEval dataset.}
\label{tab:app: energy_human}
\vspace*{-0.15in}
\end{table}

\begin{table}[h]
\centering
\resizebox{\columnwidth}{!}{  
\begin{tabular}{ccccc}
\toprule
\textbf{Methods} & \textbf{LLaMA 68M\&7B} & \textbf{Vicuna 68M\&13B} & \textbf{Deepseek 1.3\&33B} & \textbf{LLaMA-3.1 8B\&70B}\\
\midrule
SpS & 146 KJ & 250 KJ & 393 KJ & 741 KJ \\
PEARL & 193.6 KJ & 295 KJ & 428 KJ & 901 KJ \\
\rowcolor{mygreen}\textbf{SpecBranch} & \textbf{128 KJ }& \textbf{178 KJ} & \textbf{395 KJ} & \textbf{791 KJ} \\
\bottomrule
\end{tabular}
}
\caption{Energy cost of SpecBranch and baseline methods on the GSM8K dataset.}
\label{tab:app: energy_gsm9k}
\vspace*{-0.15in}
\end{table}

\subsection{Evaluation Results of Implicit Distribution}
\label{sec:appendix_result6}
As discussed in Sections~\ref{sec:analysis2} and~\ref{sec:ablation}, we primarily analyze the sensitivity of hyperparameters in the implicit methods. Here, we further examine the top-$1$ implicit distribution under different experimental setups. We conduct extensive experiments with LLaMA 68M\&7B, Deepseek 1.3B\&33B, and LLaMA-3.1 8B\&70B on HumanEval, GSM8K, and CNN/DM under various temperature settings.

\begin{figure*}[h]
\centering
\includegraphics[width=1.00\linewidth]{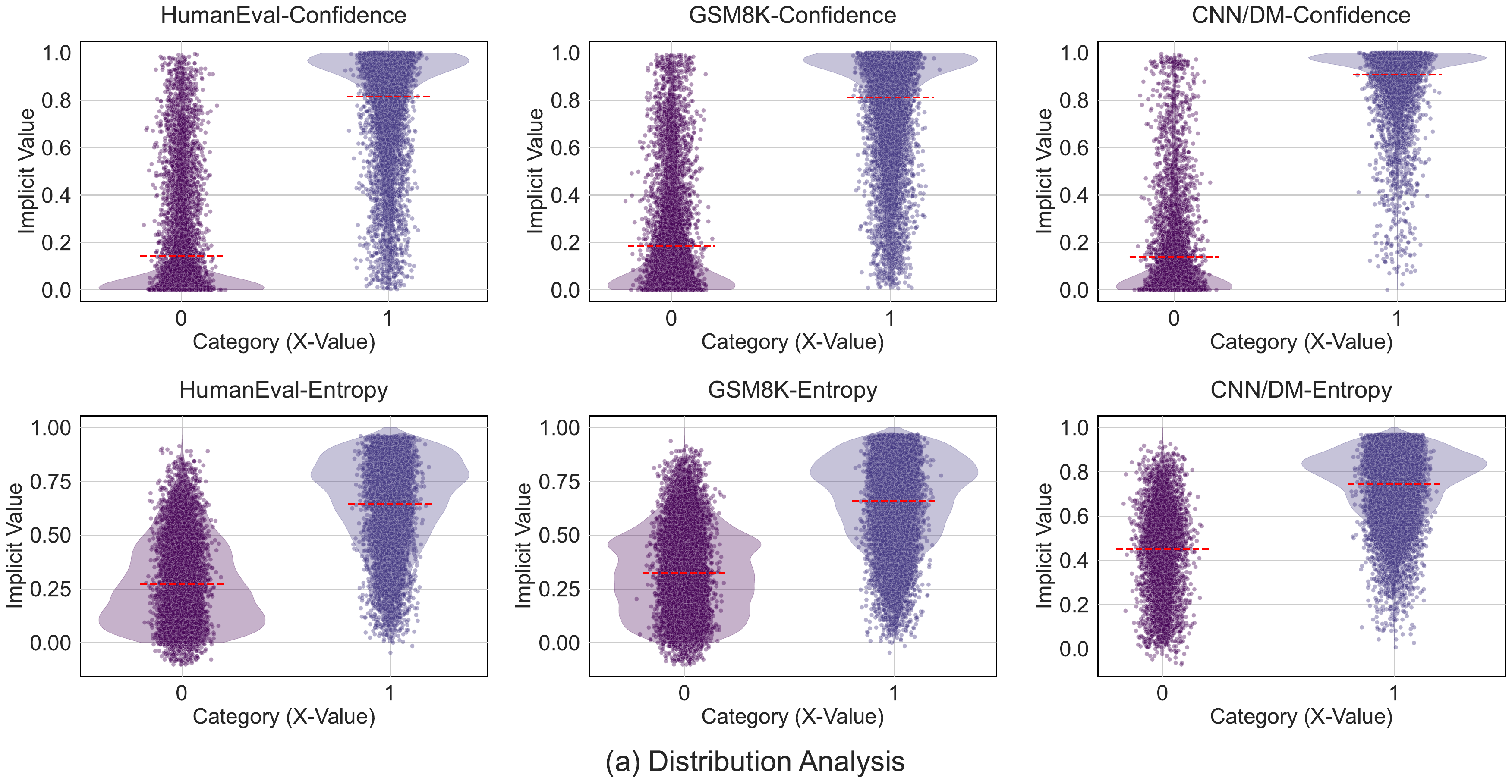}
\vspace*{-0.2in}
\caption{The top-$1$ implicit values (Confidence) distribution of LLaMA 68M\&7B on HumanEval, GSM8k and CNN/DM with $\text{temperature} = 1$. Category $0$ denotes the top-$1$ values of rejected draft tokens while $1$ denotes the corresponding values of accepted tokens.}
\vspace*{-0.15in}
\label{fig:implicit_distribution}
\end{figure*}

\textbf{Task Sensitivity} \quad 
We first explore the task sensitivity of two main implicit values: \emph{confidence} and \emph{entropy}, which measure confidence by $\max_{x_{i}} q(x_{i})$~\cite{du2024glide} and entropy as $1-\sqrt{\lambda H(x_{i})}$~\cite{agrawal2024adaedl} against pre-determined thresholds $\epsilon$. Fig.~\ref{fig:implicit_distribution} illustrates that the implicit values have different distributions across tasks. In summarization tasks (CNN/DM), both the average accepted confidence ($0.91$) and entropy ($0.77$) are significantly higher than in other tasks. Meanwhile, the rejected implicit values also vary notably ($0.26$ to $0.45$), especially for entropy in summarization tasks. This indicates that the implicit distribution is highly task-sensitive, making the selection of the stop threshold $\epsilon$ static and finding an optimal value difficult.

On the other hand, we observe that compared to entropy, confidence has a clearer and more distinct distribution for accepted and rejected tokens. This is why confidence is chosen for H-RAD, as it provides higher fidelity. Additionally, we note that around $0.5$, both confidence and entropy show considerable overlap between accepted and rejected values, which indicates a key limitation of the implicit methods.

\begin{figure*}[h]
\centering
\includegraphics[width=1.00\linewidth]{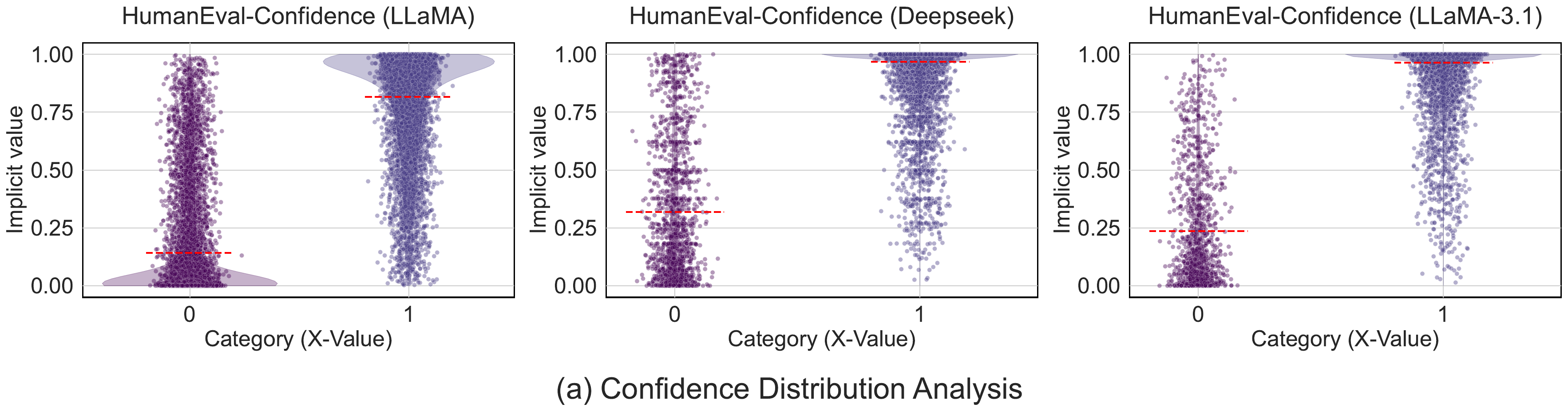}
\vspace*{-0.25in}
\caption{The top-$1$ implicit values (Confidence) distribution of LLaMA 68M\&7B, Deepseek 1.3B\&33B, and LLaMA-3,1 8B\&70B on HumanEval with $\text{temperature} = 1$. Category $0$ denotes the top-1 values of rejected draft tokens while $1$ denotes the corresponding values of accepted tokens.}
\label{fig:models_distribution}
\vspace*{-0.10in}
\end{figure*}

\textbf{Model Sensitivity} \quad
The distribution of entropy is less effective than that of confidence, so we further test the model sensitivity of implicit methods (confidence) on three different model sizes: LLaMA 68M\&7B, Deepseek 1.3B\&33B, and LLaMA-3.1 8B\&70B. As shown in Fig.~\ref{fig:models_distribution}, the average accepted confidence of better-aligned models ($0.98$) is higher than that of poorly aligned models ($0.79$), and the same holds for rejected confidence ($0.27$ against $0.11$). This demonstrates that the confidence distribution is sensitive to model pairs, with lower average confidence in poorly aligned models, resulting in a higher rate of rollback.

\begin{figure*}[h]
\centering
\includegraphics[width=1.00\linewidth]{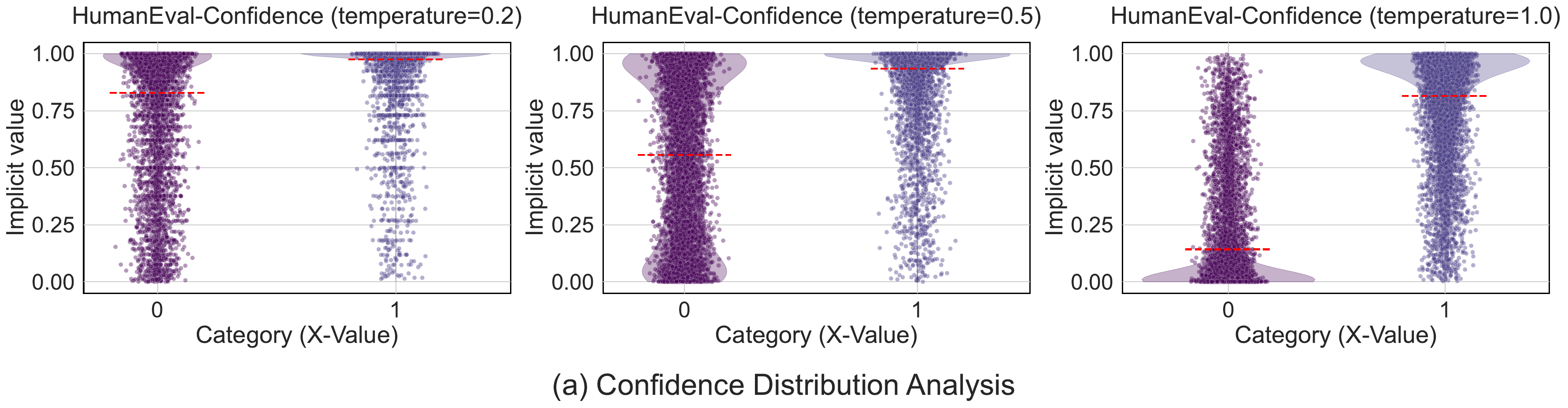}
\vspace*{-0.25in}
\caption{The top-$1$ implicit values (Confidence, Entropy) distribution of LLaMA 68M\&7B on HumanEval with $\text{temperature} = 0.2$, $0.5$ and $1$.}
\vspace*{-0.10in}
\label{fig:temperature_distribution}
\end{figure*}

\textbf{Temperature Sensitivity} \quad
Temperature plays an important role in the draft model's sampling process. Thus, we conduct additional analysis on the temperature parameter. As shown in Fig.~\ref{fig:temperature_distribution}, we observe a sharp variation in the confidence distribution with temperature, especially for rejected tokens. When $\text{temperature} = 1$, the draft model generates tokens with higher randomness, leading to a more distinct and separated confidence distribution. At $\text{temperature} = 0.7$, the average rejected confidence rises to $0.55$, overlapping more with accepted confidence. When $\text{temperature} = 0.2$, the randomness of the draft model’s sampling decreases, causing the rejected and accepted confidence distributions to overlap, making it difficult to distinguish early stop tokens. From the above, we conclude that higher temperatures increase randomness and allow the confidence distribution to better reflect whether draft tokens are accepted by the target model or not. Notably, SpecBranch uses $\text{temperature} = 1$ for top-$k$ sampling and confidence selection, optimizing the use of implicit values for better representation.

Based on the above discussion, we summarize the advantages of H-RAD over implicit methods. H-RAD significantly reduces the frequency of implicit confidence calls at the token-level by leveraging explicit target model features. This mitigates error accumulation in implicit confidence, desensitizes and reduces the dependency on thresholds. Moreover, H-RAD improves the accuracy of dynamic draft structures, resulting in a significant reduction of rollback tokens. Specifically, our H-RAD predictor is invoked only once during each draft process, incurring far lower overhead compared to the token-level implicit training predictors~\cite{huang2024specdec++}.

\section{Further Analysis and Discussion} \label{sec:appendix_futurework}

\subsection{Memory constrained scenarios}
\label{sec:appendix_memory}
We have discussed the memory consumption of SpecBranch in Section~\ref{sec:ablation}. Here, we provide a more detailed discussion about memory-constrained scenarios.

\textbf{Clarification of the Application Scenarios} \quad
First, we clarify the application scenarios of SpecBranch: most existing methods operate on a ``draft-then-verify'' sequential execution, which limits the ability to fully utilize redundant computational resources. In contrast, the main application scenarios for SpecBranch focus on environments with sufficient computational resources to enable parallel frameworks, where serialized execution is not able to adequately leverage available resources.

\textbf{SpecBranch in Memory-Abundant Scenarios.} \quad Specifically, SpecBranch is well-suited for multi-GPU parallel scenarios in cloud environments with ample GPU resources, as well as for cloud-edge collaborative settings. In such scenarios, the draft and target models are deployed on the edge and cloud devices, respectively. Additionally, it can be applied to heterogeneous processor environments, including CPU \& GPU configurations or heterogeneous GPUs. We foresee significant potential for SpecBranch in these scenarios, where both the draft and target models are independently deployed across processors. This avoids slowdown caused by memory contention. Our main experiments are conducted under these settings. Furthermore, integrating tensor parallelism (TP) with SpecBranch in these environments~\cite{pd_2024} can further enhance acceleration in the future.

\begin{table}[h]
\centering
\resizebox{\columnwidth}{!}{  
\begin{tabular}{cccccccc}
\toprule
\textbf{Methods} & \textbf{MT Bench} & \textbf{QA} & \textbf{Summarization} & \textbf{Math}  & \textbf{RAG} & \textbf{Translation} & \textbf{Avg.}\\
\midrule
Sps & 2.03$\times$ & 2.12$\times$ & 2.02$\times$ & 2.06$\times$ & 1,94$\times$ & 2.21$\times$ & 2.06$\times$ \\
SpecBranch & 3.02$\times$ & 2.87$\times$ & 2.95$\times$ & 3.21$\times$ & 2.76$\times$ & 3.28$\times$ & 3.02$\times$ \\
SpecBranch(PP) & 2.80$\times$ & 2.56$\times$ & 2.57$\times$ & 2.93$\times$ & 2.41$\times$ & 3.02$\times$ & 2.73$\times$ \\
\rowcolor{mygreen} \textbf{Performance retain} & \textbf{92.57}\% & \textbf{89.03}\% & \textbf{89.78}\% & \textbf{91.38}\% &  \textbf{87.21}\% & \textbf{91.93}\% & \textbf{89.76}\%\\
\bottomrule
\end{tabular}
}
\caption{Comparisons of Deepseek 1.3B \& 33B on the Spec-Branch tasks with the proposed SpecBench in memory-constrained scenarios.}
\label{tab:memorycons}
\vspace*{-0.14in}
\end{table}

\textbf{SpecBranch in Memory-Constrained Scenarios.} \quad We also consider memory-constrained scenarios, where resource contention between the draft and target model may arise. To mitigate this, we consider a common real-world scenario using the A100 40GB GPU. In this case, a large target model ($33$B) is deployed across two GPUs, while a smaller draft model ($1.3$B) is deployed on one of these GPUs. In such a scenario, we employ a modified pipeline parallelism (PP) version of SpecBranch to alleviate resource contention based on PEARL~\cite{liu2024parallel}. Specifically, the target model’s computation is sequential across multiple GPUs: while the target model runs on GPU $0$, the draft model can operate in parallel on GPU $1$ to generate the first $\lceil\frac{\gamma}{2}\rceil$ tokens; when the target model progresses to GPU $1$, the draft tokens generated on GPU $1$ are transferred to the idle GPU $0$, which continues generating the remaining $\lceil\frac{\gamma}{2}\rceil$ tokens, thereby avoiding memory contention. Although this PP approach introduces additional communication overhead, it effectively mitigates memory contention in this specific scenario. We conduct experiments with Deepseek 1.3B \& 33B on Spec-Bench and find that SpecBranch maintains approximately \textbf{90\%} of the reliable performance through PP, which outperforms the vanilla SD and auto-regressive decoding in memory-constrained settings. The results are shown in Table~\ref{tab:memorycons}.

Meanwhile, SpecBranch conducts extensive experiments on poorly aligned lightweight model combinations (68M \& 7B, 68M \& 13B). The experimental results demonstrate that, under H-RAD's rollback mitigation, SpecBranch exhibits better adaptability and energy efficiency for small draft models, making it more suitable for deployment in resource-constrained environments.

\begin{table}[h]
\centering
\resizebox{\columnwidth}{!}{  
\begin{tabular}{cccccccc}
\toprule
\textbf{Methods} & \textbf{MT Bench} & \textbf{QA} & \textbf{Summarization} & \textbf{Math}  & \textbf{RAG} & \textbf{Translation} & \textbf{Avg.}\\
\midrule
PEARL(Sps) & 1.74$\times$ & 1.64$\times$ & 1.70$\times$ & 1.55$\times$ & 1.56$\times$ & 1.65$\times$ & 1.64$\times$ \\
\rowcolor{mygreen} \textbf{SpecBranch \textit{w/o branch}} & \textbf{1.87$\times$} & \textbf{1.73$\times$} & \textbf{1.75$\times$} & \textbf{1.71$\times$} & \textbf{1.73$\times$} & \textbf{2.03$\times$} & \textbf{1.81$\times$} \\
\bottomrule
\end{tabular}
}
\caption{Comparisons of Vicuna 68M \& 13B on the Spec-Branch tasks with the proposed SpecBench in single GPU scenarios.}
\label{tab:singlegpu}
\vspace*{-0.14in}
\end{table}

\textbf{SpecBranch in Single GPU Scenarios.} \quad
In the extreme resource-constrained scenarios, where only a single GPU is available for inference deployment, we can still apply the pipeline parallelism (PP) strategy by offloading the draft model to the CPU (DRAM), enabling heterogeneous parallelism between the CPU and GPU. This approach is left as future work for SpecBranch. If deployment is limited to a single GPU, SpecBranch degenerates to a non-parallel framework, as discussed in Section~\ref{sec:ablation} under SpecBranch \textit{w/o branch}. In this case, the H-RAD component operates independently of the parallel framework and can be seamlessly integrated with the existing draft-then-verify methods. In contrast, PEARL's pre-verify and post-verify stages degenerate to vanilla SD under extreme resource constraints. We conduct experiments with Vicuna 68M and 13B on Spec-Bench using a single A100 GPU and find that SpecBranch \textit{w/o branch} outperforms PEARL (vanilla SD) in these single-GPU scenarios. The results are shown in Table~\ref{tab:singlegpu}.

\subsection{Discussions on Future Work.} 
Although speculative decoding has achieved remarkable success in accelerating pure-text generation, its potential across diverse downstream applications remains to be fully explored. Looking forward, we envision that Parallel Speculative Decoding (PSD) paradigms, exemplified by SpecBranch and \textsc{Double}~\cite{shen2026double}, can be seamlessly adapted into several promising avenues. \textbf{1) multimodal LLM acceleration}, where PSD can be tailored to efficiently serve large vision-language and video models through visual-semantic guidance, KV cache compression, and novel parallel decoding architectures~\cite{ kong2026parallelvlm, kong2026vision}. \textbf{2) complex agentic reasoning}, where accelerated dynamic policy exploration, and multi-perspective verification are critical for agent workflows~\cite{wu2026atlas, wu2026ssl, wu2026spark}. \textbf{3) broader efficiency optimizations}, including pipelined multi-DNN execution on heterogeneous hardware, to push the limits of modern inference systems~\cite{shen2025hetero,liu2026talon,shen2025batch,shen2025flowmesh,an2026flowhijackdynamicsawarebackdoorattack}.

\subsection{Tree Structure and Temporal mismatch}
\label{sec:appendix_future2}
In this section, we provide more detailed discussions about the tree structure and temporal mismatch for better understanding of SpecBranch (Section~\ref{sec:design}).

\begin{figure*}[h]
\centering
\includegraphics[width=0.80\linewidth]{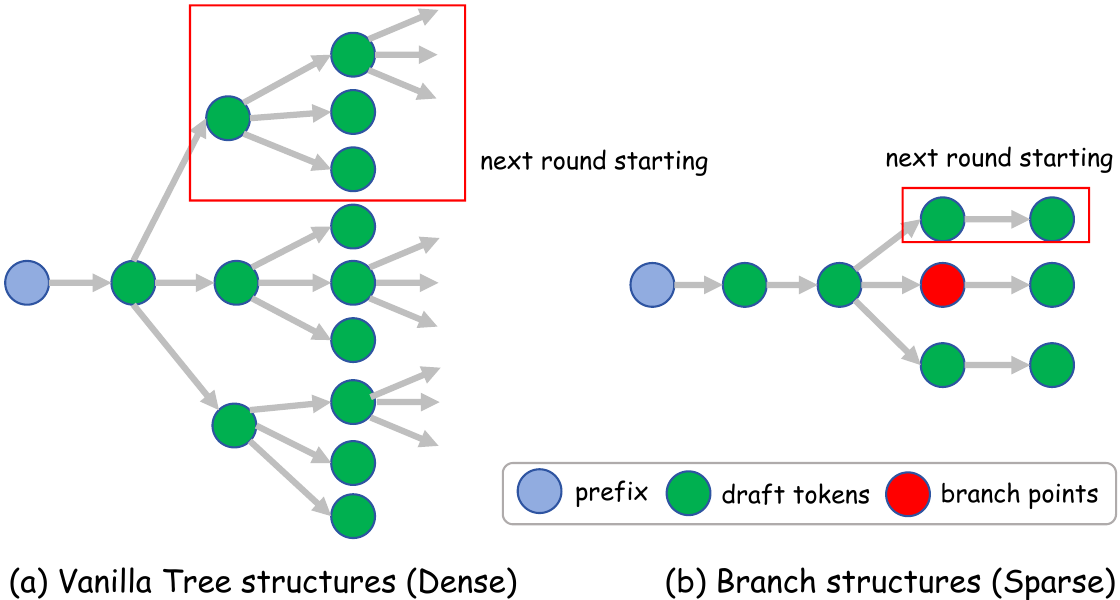}
\caption{Comparison of the vanilla tree structures (dense) and branch structures (sparse).}
\label{fig:tree}
\vspace*{-0.13in}
\end{figure*}

\textbf{Tree-based Structure}\quad
We primarily focus on the vanilla tree structures. For example, SpecInfer~\cite{miao2024specinfer} constructs a token tree using $k$ independent sequences, a topology that is constrained by the expected number of tokens it can accept, regardless of the tree size, as shown in Fig.~\ref{fig:tree}(a). However, this structure is dense, as top-$k$ sampling is applied to each token, which generates additional sequences. In the case where the draft length is $\gamma$ and the tree size is $k$, the number of tokens for each round is given by $\frac{k^{\gamma} - 1}{k - 1}$. We observe that the number of tokens in a tree structure grows exponentially with $\gamma$, causing KV-Cache storage to increase exponentially as well. To address this challenge, EAGLE2~\cite{li2024eagle2} and SEQUOIA~\cite{chen2024sequoia} employ dynamic draft tree adjustments to prune unnecessary branches, resulting in a sparse tree structure. Our SpecBranch adopts a similar sparse branch structure. Unlike traditional tree structures, where each token generates a branch, our approach, utilizing H-RAD to predict high-impact token positions, spawns sparse branch points that minimize speculative divergence and maintain parallel efficiency. As illustrated in Fig.~\ref{fig:tree}(b), we branch at $b \leq \gamma$ positions, resulting in the number of tokens per round in the branch structure being $ k \cdot \gamma + (k-1)\cdot(1-b)$, which is significantly smaller than in vanilla tree structures. In general, vanilla tree structures maintain the KV-Cache for the entire search tree, with a space complexity of \(O(k^\gamma)\), whereas SpecBranch reduces the complexity to \(O(k \cdot \gamma)\) by sparsely targeting specific branches.

Moreover, trees require complex attention mask verification to process the entire structure in parallel, whereas our approach only needs to verify the branch point. From a real-world deployment perspective, branch structures offer advantages in terms of deployment complexity and resource efficiency. Additionally, for sparse tree structures, SpecBranch can be easily integrated with them, which we leave for future work. However, a key limitation is that tree structures are not well-suited for parallel architectures, a challenge we address next.

\begin{figure*}[h]
\centering
\includegraphics[width=0.95\linewidth]{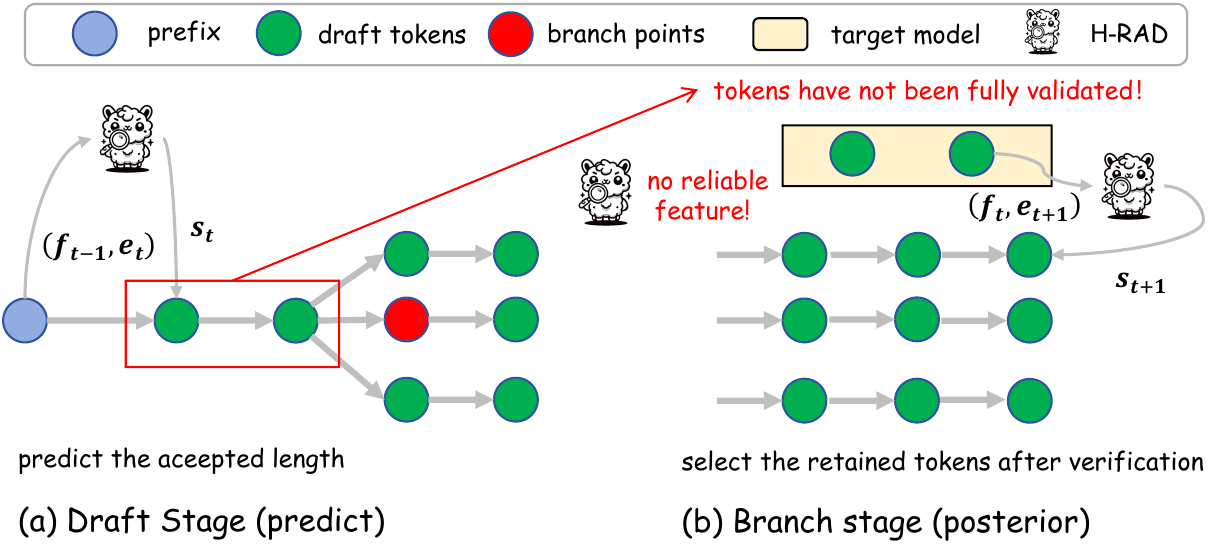}
\caption{Comparison of the Draft and Branch Stages. The parallel branch
stage introduces a temporal mismatch between drafting and verification.}
\label{fig:mismatch}
\vspace*{-0.10in}
\end{figure*}

\textbf{Temporal Mismatch}\quad
As shown in Fig.~\ref{fig:mismatch}, we observe that the parallel branch stage introduces a temporal mismatch between drafting and verification. In the draft stage (a), the previous tokens have already been verified by the target model (except for the first round, where the target model lacks feature information). As a result, H-RAD can utilize the historical target model feature pairs to predict the token generation length for the next round. In this stage, H-RAD uses features in a priori fashion for prediction. However, in the branch stage (b), the draft model proactively generates speculative branches concurrently with target model verification. This parallelization of drafting and verification causes an issue that tokens from the previous round have not been fully verified by the target model before new tokens are generated. This temporal mismatch prevents H-RAD from obtaining reliable features from the target model before new branches generate tokens.

For vanilla tree structures, temporal mismatch limits the verification process to only determining the first token of the next round. In other words, the next round starting (Fig.~\ref{fig:tree}) in tree structures is not just one token, but rather $(\gamma - 1) \cdot k$ tokens from a partial tree. This means that during the parallel phase, the number of tokens in a dense tree structure will reach $(\gamma - 1) \cdot k \cdot \frac{k^{\gamma} - 1}{k - 1}$, which is $(\gamma - 1) \cdot k$ times higher than the KV-Cache in the previous intermediate steps, exacerbating memory overhead. On the other hand, branch structures, due to their sparsity, only retain one branch after each verification step. This ensures that each new round starts afresh and mitigates the memory overhead.

\textbf{Posterior Approach}\quad
To address this temporal mismatch, we introduce a posterior approach in Section~\ref{sec:Posterior Drafting}. Specifically, we wait until the previous tokens are fully verified by the target model. Since $\gamma$ represents the speed ratio $c$ between the draft and target models, by the time this happens, the branch has also completed token generation. Then, for the remaining branch $V$, we use the features $(f_t, e_{t+1})$ from the current round as input to H-RAD and select the retained tokens $\mathcal{H}_t$ after the verification step. This posterior approach effectively resolves the temporal mismatch between verification and drafting, ensuring that H-RAD always uses the most up-to-date and relevant context. It also leverages parallelism, making the time loss from the posterior approach negligible. However, on the other hand, unlike in the draft stage, we cannot implement early stopping. While the parallelization mostly mitigates time impact due to bottlenecks at the target model, this introduces additional token waste. Moreover, this feature utilization method does not fully align with the current mainstream methods such as EAGLE~\cite{li2024eagle}.

\begin{wrapfigure}{r}{0.35\textwidth}
    \begin{center}
    \includegraphics[width=0.35\columnwidth]{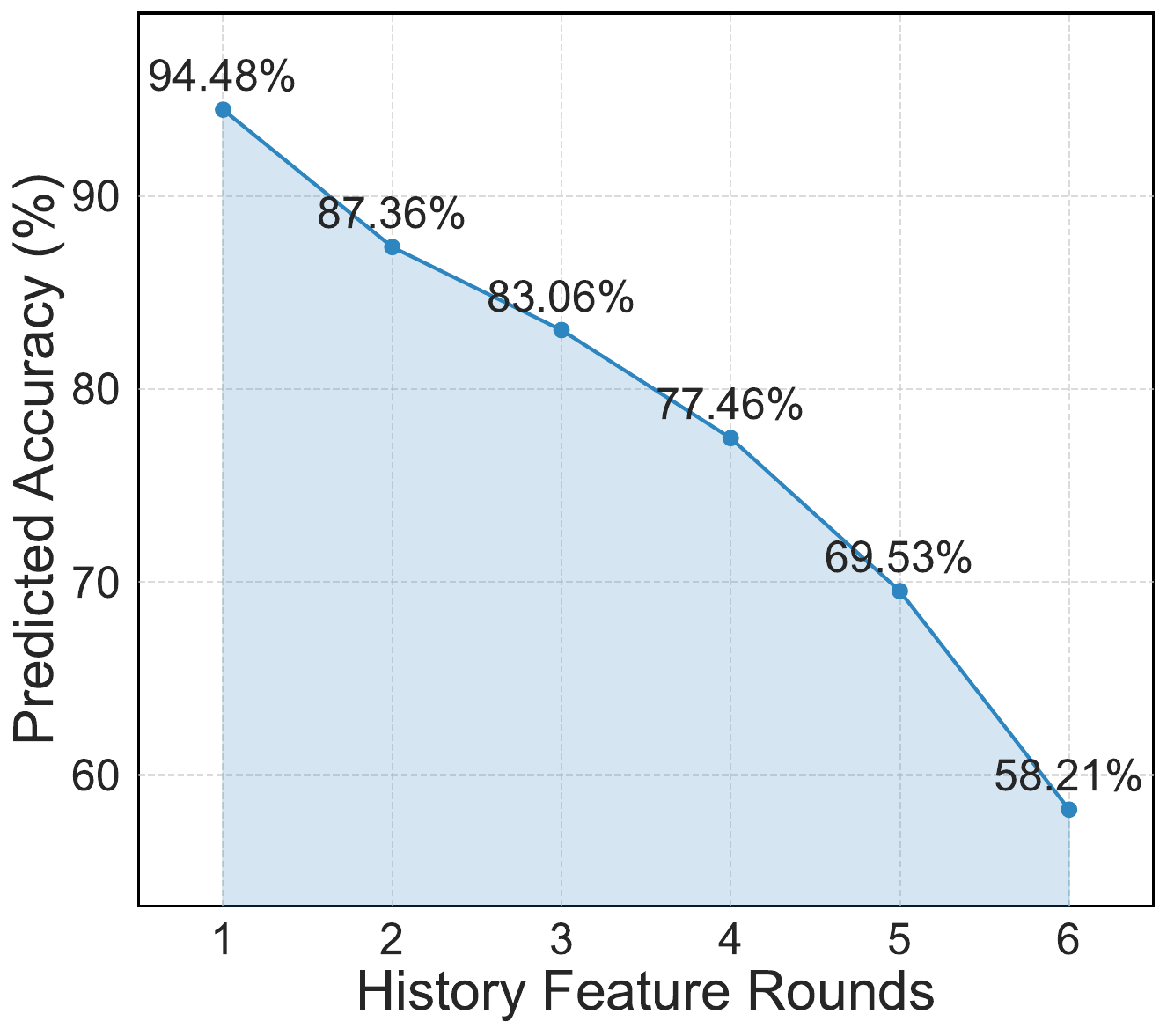}
    \vspace{-0.6cm}
    \caption{{Predictive capability of the features progressively decay of LLaMA 68M\&7B on HumanEval.}}
    \label{fig:history}
    \end{center}
    \vspace{-0.1in}
\end{wrapfigure}
\textbf{A priori Approach} \quad We further explore a priori methods to unify the verification processes between the draft and branch stages. We leverage the \emph{temporal locality} of transformer hidden states, where features from earlier time steps, though slightly outdated, still retain sufficient predictive power. Thus, H-RAD can proactively reuse features from the target model, specifically $(f_{t-1}, e_{t})$, to predict $\mathcal{H}_t$, whereas the posterior approach uses $(f_t, e_{t+1})$ as input. In other words, if we cannot reuse features from the previous round's token, we rely on token features from the two rounds back, which exhibits a temporal decay, where the predictive capability of the target model features diminishes across multiple rounds. 

To quantify such decay, we conduct a small test of LLaMA 68M\&7B on HumanEval as shown in Fig.~\ref{fig:history}. It measures the prediction accuracy of the MLP used in H-RAD. We observe that, as the distance and number of rounds increase -- transitioning from $(f_t, e_{t+1})$ to $(f_{t-n}, e_{t+1-n})$ -- the contextual information and predictive capability of the features decay gradually. However, we note that the features from the previous two rounds, $(f_{t-1}, e_{t})$, still exhibit some predictive capacity. In scenarios where speed is not as critical, these stale features may serve as a viable surrogate for the more recent features $(f_t, e_{t+1})$. This a priori approach better unifies the operations of the draft and branch stages. While the acceleration effect is not as significant as the posterior approach, the early stopping mechanism reduces the number of tokens generated. Additionally, this method can be combined with EAGLE~\cite{li2024eagle} to reuse features with the draft model. In future work, we aim to explore the integration of EAGLE with parallel mechanisms to contribute to the diversity of the SD community. 

\subsection{Group Speculative Decoding and Heterogeneous Devices}
\label{sec:appendix_future3}

\textbf{Group SD within a Single GPU Cluster} \quad
In SpecBranch, we rigorously validate the acceleration effectiveness for a single draft-target model pair. However, as Model-as-a-Service (MaaS) becomes more prevalent, serving multiple model instances within a cluster is increasingly important for maximizing resource utilization. We plan to extend our system to support simultaneous SD for multiple draft-target pairs with more than one draft model per GPU and high throughput. For example, in an 8\(\times\)A100 80G configuration, we can deploy $7$ different target models ($33$B) on $7$ GPUs and up to $15$ draft models ($1.3$B) on the remaining GPUs. These draft models can be paired with the same or different target models through advanced communication and network design.

\textbf{Speculative Decoding on Heterogeneous Devices} \quad
During validation, we find that large models often require multiple high-end GPUs to fit into memory (e.g., a $70$B-parameter model experiment requires at least $4$\(\times\)A100 40G GPUs). While one solution discussed in the previous section is to pack multiple draft models in a single device, the prerequisite for large-memory devices remains unchanged. However, we discover that under the setting of SD, homogeneous deployment is unnecessary, i.e., the draft models do not need to be co-located on the same device as the target models. Thus, we present a heterogeneous structure that runs smaller draft models on consumer-grade GPUs (e.g., RTX 3090) and connects them over the network to target models on data-center GPUs (e.g., A100). The challenges on the system level, including synchronization, communication latency, and workload balancing, remain as future work for SpecBranch.

\label{sec:appendix_future4}

\newpage

\end{document}

%% file: pseudo.tex
\begin{algorithm}[H]
\SetAlgoLined
\definecolor{mygreen}{RGB}{79, 156, 94}
\DeclareRobustCommand{\cmtsty}[1]{\itshape\color{mygreen}#1}
\SetCommentSty{cmtsty}
\DontPrintSemicolon
\newcommand{\cmark}{\ding{51}}  
\newcommand{\xmark}{\ding{55}}  

\caption{Algorithm of SpecBranch}
\KwIn{
  prefix $I$,           
  draft model $M_q$,         
  target model $M_p$,        
  max length $L$,            
  target model feature $f$,
  target model embedding $e$,
  predicted output $s_t$,
  confidence threshold $\epsilon$,  
  num branches $B$,          
  max gamma $\gamma_m$,      
  execution mode $E$
}
\KwOut{generated sequence $\mathbf{x}$}

$\mathbf{x} \leftarrow I$,
$\gamma \leftarrow \gamma_m$,
$E \leftarrow \texttt{DRAFTING}$\;

\While{$|\mathbf{x}| < L$}{
  \uIf{$E = \texttt{DRAFTING}$}{
    \tcp{Drafting phase: only drafting candidate tokens}
    $\gamma = \texttt{Predictor}(f_{t-1}, e_t, \epsilon, \gamma_m)$ \; 
    
    \For{$i \leftarrow 1 \KwTo \gamma$}{
      $q_i \leftarrow M_q\bigl(\mathbf{x} + [x_1,\dots,x_{i-1}]\bigr)$ \;
      $x_i \sim q_i$ \;
    }
    $E \leftarrow \texttt{VERIFICATION}$ \;
  }{
    \tcp{Verification phase: evaluate branches}
    \ForEach{$b \in \{1,\dots,B\}$}{
      \For{$i \leftarrow \gamma+1 \KwTo 2\gamma$}{
        $q_{(b,i)} \leftarrow \texttt{Mask}(M_q\bigl(\mathbf{x} + [x_{(b,1)},\dots,x_{(b,i-1)}]\bigr))$\;
        $x_{(b,i)} \sim q_{(b,i)}$\;
        \uIf{$q_{(b,i)}[x_{(b,i)}] < \epsilon$}{
          \textbf{keep the invalid position}}\;
        }
      }
    }
    
    $(p_1,\dots,p_\gamma) \leftarrow \bigl(M_p(\mathbf{x} + [x_1]),\dots,M_p(\mathbf{x} + [x_1,\dots,x_\gamma])\bigr)$ \;

    Retrieve $(q_1,\dots,q_\gamma)$ from cache \;

    \For{$i \leftarrow 1 \KwTo \gamma$}{
      $r_i \sim U(0,1)$\;
    }

    $n \leftarrow \min\bigl(\{\,i-1 \mid r_i > \frac{p_i[x_i]}{q_i[x_i]}\}\cup\{\gamma\}\bigr)$ \;

    \uIf{$n = \gamma$}{
      \tcp{All drafted tokens accepted}
      \uIf{$\exists\,b: r_b \le \frac{p[x_{\!n+1}]}{q_{(b,n+1)}[x_{\!n+1}]}$}{
        $b^* \leftarrow \arg\max_{b}\,r_b$\;
        $\mathbf{x} \leftarrow \mathbf{x} + [\,x_{(b^*,n+1)},\ldots,x_{(b^*,\gamma)}]$\;
        $\gamma \leftarrow \texttt{Predictor}(f_{t}, e_{t+1}, \epsilon,  \gamma_m)$\;
        $E \leftarrow \texttt{VERIFICATION}$\;
      }
      \Else{
        \tcp{\xmark\ Reject next token — fallback to target}
        $y \sim \mathcal{N}\bigl(\max(0,\,p_{n+1} - q_{(b,n+1)})\bigr)$\;
        $\mathbf{x} \leftarrow \mathbf{x} + [y]$\;
        $E \leftarrow \texttt{DRAFTING}$\;
      }
    }
    \Else{
      \tcp{Rejection occurred at position $n$}
      $y \sim \mathcal{N}\bigl(\max(0,\,p_{n+1} - q_{n+1})\bigr)$\;
      $\mathbf{x} \leftarrow \mathbf{x} + [\,x_1,\dots,x_n,\,y\,]$\;
      $E \leftarrow \texttt{DRAFTING}$\;
    }
}
\Return \texttt{x}
\end{algorithm}

%% file: reference.bib
@article{chen2024sequoia,
  title={Sequoia: Scalable and robust speculative decoding},
  author={Chen, Zhuoming and May, Avner and Svirschevski, Ruslan and Huang, Yu-Hsun and Ryabinin, Max and Jia, Zhihao and Chen, Beidi},
  journal={Advances in Neural Information Processing Systems},
  volume={37},
  pages={129531--129563},
  year={2024}
}

@article{achiam2023gpt,
  title={Gpt-4 technical report},
  author={Achiam, Josh and Adler, Steven and Agarwal, Sandhini and Ahmad, Lama and Akkaya, Ilge and Aleman, Florencia Leoni and Almeida, Diogo and Altenschmidt, Janko and Altman, Sam and Anadkat, Shyamal and others},
  journal={arXiv preprint arXiv:2303.08774},
  year={2023}
}

@inproceedings{miao2024specinfer,
  title={Specinfer: Accelerating large language model serving with tree-based speculative inference and verification},
  author={Miao, Xupeng and Oliaro, Gabriele and Zhang, Zhihao and Cheng, Xinhao and Wang, Zeyu and Zhang, Zhengxin and Wong, Rae Ying Yee and Zhu, Alan and Yang, Lijie and Shi, Xiaoxiang and others},
  booktitle={Proceedings of the 29th ACM International Conference on Architectural Support for Programming Languages and Operating Systems, Volume 3},
  pages={932--949},
  year={2024}
}

@article{team2023gemini,
  title={Gemini: a family of highly capable multimodal models},
  author={Team, Gemini and Anil, Rohan and Borgeaud, Sebastian and Alayrac, Jean-Baptiste and Yu, Jiahui and Soricut, Radu and Schalkwyk, Johan and Dai, Andrew M and Hauth, Anja and Millican, Katie and others},
  journal={arXiv preprint arXiv:2312.11805},
  year={2023}
}

@article{guo2025deepseek,
  title={Deepseek-r1: Incentivizing reasoning capability in llms via reinforcement learning},
  author={Guo, Daya and Yang, Dejian and Zhang, Haowei and Song, Junxiao and Zhang, Ruoyu and Xu, Runxin and Zhu, Qihao and Ma, Shirong and Wang, Peiyi and Bi, Xiao and others},
  journal={arXiv preprint arXiv:2501.12948},
  year={2025}
}

@article{bai2023qwen,
  title={Qwen technical report},
  author={Bai, Jinze and Bai, Shuai and Chu, Yunfei and Cui, Zeyu and Dang, Kai and Deng, Xiaodong and Fan, Yang and Ge, Wenbin and Han, Yu and Huang, Fei and others},
  journal={arXiv preprint arXiv:2309.16609},
  year={2023}
}

@inproceedings{jimenez2001dynamic,
  title={Dynamic branch prediction with perceptrons},
  author={Jim{\'e}nez, Daniel A and Lin, Calvin},
  booktitle={Proceedings HPCA Seventh International Symposium on High-Performance Computer Architecture},
  pages={197--206},
  year={2001},
  organization={IEEE}
}

@inproceedings{shi2019applying,
  title={Applying deep learning to the cache replacement problem},
  author={Shi, Zhan and Huang, Xiangru and Jain, Akanksha and Lin, Calvin},
  booktitle={Proceedings of the 52nd Annual IEEE/ACM International Symposium on Microarchitecture},
  pages={413--425},
  year={2019}
}

@article{agrawal2024adaedl,
  title={AdaEDL: Early Draft Stopping for Speculative Decoding of Large Language Models via an Entropy-based Lower Bound on Token Acceptance Probability},
  author={Agrawal, Sudhanshu and Jeon, Wonseok and Lee, Mingu},
  journal={arXiv preprint arXiv:2410.18351},
  year={2024}
}

@inproceedings{narayanan2019pipedream,
  title={PipeDream: Generalized pipeline parallelism for DNN training},
  author={Narayanan, Deepak and Harlap, Aaron and Phanishayee, Amar and Seshadri, Vivek and Devanur, Nikhil R and Ganger, Gregory R and Gibbons, Phillip B and Zaharia, Matei},
  booktitle={Proceedings of the 27th ACM symposium on operating systems principles},
  pages={1--15},
  year={2019}
}

@article{cai2024medusa,
  title={Medusa: Simple llm inference acceleration framework with multiple decoding heads},
  author={Cai, Tianle and Li, Yuhong and Geng, Zhengyang and Peng, Hongwu and Lee, Jason D and Chen, Deming and Dao, Tri},
  journal={arXiv preprint arXiv:2401.10774},
  year={2024}
}

@article{liu2024kangaroo,
  title={Kangaroo: Lossless self-speculative decoding for accelerating llms via double early exiting},
  author={Liu, Fangcheng and Tang, Yehui and Liu, Zhenhua and Ni, Yunsheng and Tang, Duyu and Han, Kai and Wang, Yunhe},
  journal={Advances in Neural Information Processing Systems},
  volume={37},
  pages={11946--11965},
  year={2024}
}

@article{li2024eagle,
  title={Eagle: Speculative sampling requires rethinking feature uncertainty},
  author={Li, Yuhui and Wei, Fangyun and Zhang, Chao and Zhang, Hongyang},
  journal={arXiv preprint arXiv:2401.15077},
  year={2024}
}

@article{du2024glide,
  title={Glide with a cape: A low-hassle method to accelerate speculative decoding},
  author={Du, Cunxiao and Jiang, Jing and Yuanchen, Xu and Wu, Jiawei and Yu, Sicheng and Li, Yongqi and Li, Shenggui and Xu, Kai and Nie, Liqiang and Tu, Zhaopeng and others},
  journal={arXiv preprint arXiv:2402.02082},
  year={2024}
}

@article{li2024eagle2,
  title={Eagle-2: Faster inference of language models with dynamic draft trees},
  author={Li, Yuhui and Wei, Fangyun and Zhang, Chao and Zhang, Hongyang},
  journal={arXiv preprint arXiv:2406.16858},
  year={2024}
}

@article{zhao2024ouroboros,
  title={Ouroboros: Generating Longer Drafts Phrase by Phrase for Faster Speculative Decoding},
  author={Zhao, Weilin and Huang, Yuxiang and Han, Xu and Xu, Wang and Xiao, Chaojun and Zhang, Xinrong and Fang, Yewei and Zhang, Kaihuo and Liu, Zhiyuan and Sun, Maosong},
  journal={arXiv preprint arXiv:2402.13720},
  year={2024}
}

@article{loshchilov2017decoupled,
  title={Decoupled weight decay regularization},
  author={Loshchilov, Ilya and Hutter, Frank},
  journal={arXiv preprint arXiv:1711.05101},
  year={2017}
}

@article{zhang2024adaeagle,
  title={AdaEAGLE: Optimizing Speculative Decoding via Explicit Modeling of Adaptive Draft Structures},
  author={Zhang, Situo and Wang, Hankun and Ma, Da and Zhu, Zichen and Chen, Lu and Lan, Kunyao and Yu, Kai},
  journal={arXiv preprint arXiv:2412.18910},
  year={2024}
}

@article{liu2024parallel,
  title={Parallel speculative decoding with adaptive draft length},
  author={Liu, Tianyu and Li, Yun and Lv, Qitan and Liu, Kai and Zhu, Jianchen and Hu, Winston},
  journal={arXiv preprint arXiv:2408.11850},
  year={2024}
}

@article{fu2024break,
  title={Break the sequential dependency of llm inference using lookahead decoding},
  author={Fu, Yichao and Bailis, Peter and Stoica, Ion and Zhang, Hao},
  journal={arXiv preprint arXiv:2402.02057},
  year={2024}
}

@article{chen2023accelerating,
  title={Accelerating large language model decoding with speculative sampling},
  author={Chen, Charlie and Borgeaud, Sebastian and Irving, Geoffrey and Lespiau, Jean-Baptiste and Sifre, Laurent and Jumper, John},
  journal={arXiv preprint arXiv:2302.01318},
  year={2023}
}

@inproceedings{leviathan2023fast,
  title={Fast inference from transformers via speculative decoding},
  author={Leviathan, Yaniv and Kalman, Matan and Matias, Yossi},
  booktitle={International Conference on Machine Learning},
  pages={19274--19286},
  year={2023},
  organization={PMLR}
}

@article{stern2018blockwise,
  title={Blockwise parallel decoding for deep autoregressive models},
  author={Stern, Mitchell and Shazeer, Noam and Uszkoreit, Jakob},
  journal={Advances in Neural Information Processing Systems},
  volume={31},
  year={2018}
}

@article{brown2020language,
  title={Language models are few-shot learners},
  author={Brown, Tom and Mann, Benjamin and Ryder, Nick and Subbiah, Melanie and Kaplan, Jared D and Dhariwal, Prafulla and Neelakantan, Arvind and Shyam, Pranav and Sastry, Girish and Askell, Amanda and others},
  journal={Advances in neural information processing systems},
  volume={33},
  pages={1877--1901},
  year={2020}
}

@article{mamou2024dynamic,
  title={Dynamic speculation lookahead accelerates speculative decoding of large language models},
  author={Mamou, Jonathan and Pereg, Oren and Korat, Daniel and Berchansky, Moshe and Timor, Nadav and Wasserblat, Moshe and Schwartz, Roy},
  journal={arXiv preprint arXiv:2405.04304},
  year={2024}
}

@article{zhang2023draft,
  title={Draft \& verify: Lossless large language model acceleration via self-speculative decoding},
  author={Zhang, Jun and Wang, Jue and Li, Huan and Shou, Lidan and Chen, Ke and Chen, Gang and Mehrotra, Sharad},
  journal={arXiv preprint arXiv:2309.08168},
  year={2023}
}

@article{huang2024specdec++,
  title={Specdec++: Boosting speculative decoding via adaptive candidate lengths},
  author={Huang, Kaixuan and Guo, Xudong and Wang, Mengdi},
  journal={arXiv preprint arXiv:2405.19715},
  year={2024}
}

@inproceedings{Nallapati_Zhou_dos,   
title={Abstractive Text Summarization Using Sequence-to-Sequence RNNs and Beyond},  url={http://dx.doi.org/10.18653/v1/k16-1028},  DOI={10.18653/v1/k16-1028},  booktitle={Proceedings of The 20th SIGNLL Conference on Computational Natural Language Learning},  author={Nallapati, Ramesh and Zhou, Bowen and dos Santos, Cicero and Gulcehre, Caglar and Xiang, Bing},  year={2016},  month={Jan},  language={en-US}  }

@article{Cobbe_Kosaraju_Bavarian_Hilton_Nakano_Hesse_Schulman_2021,   
title={Training Verifiers to Solve Math Word Problems},  journal={Cornell University - arXiv,Cornell University - arXiv},  author={Cobbe, Karl and Kosaraju, Vineet and Bavarian, Mohammad and Hilton, Jacob and Nakano, Reiichiro and Hesse, Christopher and Schulman, John},  year={2021},  month={Oct},  language={en-US}  }

@article{chen2021codex,
  title={Evaluating Large Language Models Trained on Code},
  author={Mark Chen and Jerry Tworek and Heewoo Jun and Qiming Yuan and Henrique Ponde de Oliveira Pinto and Jared Kaplan and Harri Edwards and Yuri Burda and Nicholas Joseph and Greg Brockman et al.},
  year={2021},
  eprint={2107.03374},
  archivePrefix={arXiv},
  primaryClass={cs.LG}
}

@article{xia2024unlocking,
  title={Unlocking Efficiency in Large Language Model Inference: A Comprehensive Survey of Speculative Decoding},
  author={Heming Xia and Zhe Yang and Qingxiu Dong and Peiyi Wang and Yongqi Li and Tao Ge and Tianyu Liu and Wenjie Li and Zhifang Sui},
  year={2024},
  month={Jan},
  language={en-US}
}

@article{yang2024multi,
  title={Multi-Candidate Speculative Decoding},
  author={Sen Yang and Shujian Huang and Xinyu Dai and Jiajun Chen},
  year={2024},
  month={Jan},
  language={en-US}
}

@article{zheng2023judging,
  title={Judging LLM-as-a-judge with MT-Bench and Chatbot Arena},
  author={Lianmin Zheng and Wei-Lin Chiang and Ying Sheng and Siyuan Zhuang and Zhanghao Wu and Yonghao Zhuang and Zi Lin and Zhuohan Li and Dacheng Li and Eric.P Xing and Hao Zhang and JosephE. Gonzalez and Ion Stoica},
  year={2023},
  month={Jun},
  language={en-US}
}

@misc{deepseek-coder,
  author={Daya Guo and Qihao Zhu and Dejian Yang and Zhenda Xie and Kai Dong and Wentao Zhang and Guanting Chen and Xiao Bi and Y. Wu and Y.K. Li and Fuli Luo and Yingfei Xiong and Wenfeng Liang},
  title={DeepSeek-Coder: When the Large Language Model Meets Programming -- The Rise of Code Intelligence},
  journal={CoRR},
  volume={abs/2401.14196},
  year={2024},
  url={https://arxiv.org/abs/2401.14196}
}

@misc{grattafiori2024llama3herdmodels,
  title={The Llama 3 Herd of Models},
  author={Aaron Grattafiori and Abhimanyu Dubey and Abhinav Jauhri and Abhinav Pandey and Abhishek Kadian and Ahmad Al-Dahle and Aiesha Letman and Akhil Mathur and Alan Schelten and Alex Vaughan and Amy Yang and Angela Fan and Anirudh Goyal et al.},
  year={2024},
  eprint={2407.21783},
  archivePrefix={arXiv},
  primaryClass={cs.AI},
  url={https://arxiv.org/abs/2407.21783}
}

@article{touvron2023llama,
  title={LLaMA: Open and Efficient Foundation Language Models},
  author={Hugo Touvron and Thibaut Lavril and Gautier Izacard and Xavier Martinet and Marie-Anne Lachaux and Timoth’ee Lacroix and Baptiste Rozière and Naman Goyal and Eric Hambro and Faisal Azhar and Aurelien Rodriguez and Armand Joulin and Edouard Grave and Guillaume Lample},
  year={2023},
  language={en-US}
}

@article{pd_2024,   title={DistServe: Disaggregating Prefill and Decoding for Goodput-optimized Large Language Model Serving},  author={Zhong, Yinmin and Liu, Shengyu and Chen, Junda and Hu, Jianbo and Zhu, Yibo and Liu, Xuanzhe and Jin, Xin and Zhang, Hao},  year={2024},  month={Jan},  language={en-US}  }

@article{chawla2002smote,
  title={SMOTE: synthetic minority over-sampling technique},
  author={Chawla, Nitesh V and Bowyer, Kevin W and Hall, Lawrence O and Kegelmeyer, W Philip},
  journal={Journal of artificial intelligence research},
  volume={16},
  pages={321--357},
  year={2002}
}

@article{timor2024distributed,
  title={Distributed speculative inference (dsi): Speculation parallelism for provably faster lossless language model inference},
  author={Timor, Nadav and Mamou, Jonathan and Korat, Daniel and Berchansky, Moshe and Pereg, Oren and Wasserblat, Moshe and Galanti, Tomer and Gordon, Michal and Harel, David},
  journal={arXiv preprint arXiv:2405.14105},
  year={2024}
}

@inproceedings{shen2025hetero,
  title={Hetero 2 Pipe: Pipelining Multi-DNN Inference on Heterogeneous Mobile Processors under Co-Execution Slowdown},
  author={Shen, Yuhao and Wang, Zichen and Wang, Tianyi and Gu, Chaojie and Wen, Zhenyu and Shu, Yuanchao and Wang, Cong},
  booktitle={2025 IEEE 45th International Conference on Distributed Computing Systems (ICDCS)},
  pages={483--493},
  year={2025},
  organization={IEEE}
}

@article{kong2026parallelvlm,
  title={ParallelVLM: Lossless Video-LLM Acceleration with Visual Alignment Aware Parallel Speculative Decoding},
  author={Kong, Quan and Shen, Yuhao and Ji, Yicheng and Li, Huan and Wang, Cong},
  journal={arXiv preprint arXiv:2603.19610},
  year={2026}
}

@article{kong2026vision,
  title={Vision-TTT: Efficient and Expressive Visual Representation Learning with Test-Time Training},
  author={Kong, Quan and Xiao, Yanru and Shen, Yuhao and Wang, Cong},
  journal={arXiv preprint arXiv:2603.00518},
  year={2026}
}

@article{wu2026atlas,
  title={Atlas: Orchestrating Heterogeneous Models and Tools for Multi-Domain Complex Reasoning},
  author={Wu, Jinyang and Zhai, Guocheng and Jin, Ruihan and Yuan, Jiahao and Shen, Yuhao and Zhang, Shuai and Wen, Zhengqi and Tao, Jianhua},
  journal={arXiv preprint arXiv:2601.03872},
  year={2026}
}

@article{wu2026ssl,
  title={SSL: Sweet Spot Learning for Differentiated Guidance in Agentic Optimization},
  author={Wu, Jinyang and Yang, Changpeng and Shen, Yuhao and Xu, Fangzhi and Ni, Bolin and Liao, Chonghua and Liu, Yuchen and Wang, Hongzhen and Nie, Shuai and Zhang, Shuai and others},
  journal={arXiv preprint arXiv:2601.22491},
  year={2026}
}

@article{wu2026spark,
  title={Spark: Strategic Policy-Aware Exploration via Dynamic Branching for Long-Horizon Agentic Learning},
  author={Wu, Jinyang and Yang, Shuo and Yang, Changpeng and Shen, Yuhao and Zhang, Shuai and Wen, Zhengqi and Tao, Jianhua},
  journal={arXiv preprint arXiv:2601.20209},
  year={2026}
}

@article{liu2026talon,
  title={TALON: Confidence-Aware Speculative Decoding with Adaptive Token Trees},
  author={Liu, Tianyu and Lv, Qitan and Shen, Yuhao and Sun, Xiao and Sun, Xiaoyan},
  journal={arXiv preprint arXiv:2601.07353},
  year={2026}
}

@article{shen2026double,
  title={Double: Breaking the Acceleration Limit via Double Retrieval Speculative Parallelism},
  author={Shen, Yuhao and Liu, Tianyu and Shen, Junyi and Wu, Jinyang and Kong, Quan and Huan, Li and Wang, Cong},
  journal={arXiv preprint arXiv:2601.05524},
  year={2026}
}

@article{shen2025batch,
  title={Batch Query Processing and Optimization for Agentic Workflows},
  author={Shen, Junyi and Wadlom, Noppanat and Lu, Yao},
  journal={arXiv preprint arXiv:2509.02121},
  year={2025}
}

@article{shen2025flowmesh,
  title={FlowMesh: A Service Fabric for Composable LLM Workflows},
  author={Shen, Junyi and Wadlom, Noppanat and Zhou, Lingfeng and Wang, Dequan and Miao, Xu and Fang, Lei and Lu, Yao},
  journal={arXiv preprint arXiv:2510.26913},
  year={2025}
}

@misc{an2026flowhijackdynamicsawarebackdoorattack,
      title={FlowHijack: A Dynamics-Aware Backdoor Attack on Flow-Matching Vision-Language-Action Models}, 
      author={Xinyuan An and Tao Luo and Gengyun Peng and Yaobing Wang and Kui Ren and Dongxia Wang},
      year={2026},
      eprint={2604.09651},
      archivePrefix={arXiv},
      primaryClass={cs.CV},
      url={https://arxiv.org/abs/2604.09651}, 
}
